\documentclass[useAMS,usenatbib]{mn2e}
\usepackage{graphicx}
\usepackage{rotating}
\usepackage{pdflscape}

\title[Origin and Loss of nebula-captured hydrogen envelopes]
  {Origin and Loss of nebula-captured hydrogen envelopes\\ from `sub'- to `super-Earths' in the habitable zone\\ of Sun-like stars}
\author[H. Lammer et al.]
  {H.~Lammer,$^1$ A. St\"{okl},$^2$ N. V.~Erkaev,$^{3,4}$ E. A. Dorfi,$^2$ P. Odert,$^{1,5}$ M. G\"{u}del,$^2$
  \newauthor Yu. N. Kulikov,$^6$ K. G.~Kislyakova,$^{1}$ M. Leitzinger$^5$
 \\
  $^1$Space Research Institute, Austrian Academy of Sciences, Schmiedlstr. 6, A-8042, Graz, Austria\\
  $^2$Institute for Astronomy, University of Vienna,
T\"{u}rkenschanzstrasse 17, 1180 Vienna, Austria\\
  $^3$Institute for Computational Modelling, Russian Academy of Sciences,
    Krasnoyarsk 36, Russian Federation\\
$^4$Siberian Federal University, Krasnoyarsk, Russian Federation\\
  $^5$Institute of Physics, University of Graz, Universit\"{a}tsplatz 5, A-8010 Graz, Austria\\
$^6$Polar Geophysical Institute, Russian Academy of Sciences, Khalturina Str. 15,
Murmansk 183010, Russian Federation}
\date{Released 2013 Xxxxx XX}

\pagerange{\pageref{firstpage}--\pageref{lastpage}} \pubyear{2012}

\def\LaTeX{L\kern-.36em\raise.3ex\hbox{a}\kern-.15em
    T\kern-.1667em\lower.7ex\hbox{E}\kern-.125emX}

\begin{document}

\label{firstpage}

\maketitle

\begin{abstract}
We investigate the origin and loss of captured hydrogen envelopes from
protoplanets having masses in a range between
`sub-Earth'-like bodies of 0.1$M_{\oplus}$ and `super-Earths' with
5$M_{\oplus}$ in the habitable zone at 1 AU of a Sun like G star,
assuming that their rocky cores had formed before the nebula gas
dissipated. We model the gravitational attraction and accumulation
of nebula gas around a planet's core as a function of
protoplanetary luminosity during accretion and calculate the
resulting surface temperature by solving the hydrostatic structure
equations for the protoplanetary nebula. Depending on nebular
properties, such as the dust grain depletion factor, planetesimal
accretion rates, and resulting luminosities, for planetary bodies
of 0.1--1$M_{\oplus}$ we obtain hydrogen envelopes with masses
between $\sim 2.5\times 10^{19}$--$1.5\times 10^{26}$ g. For
`super-Earths' with masses between 2--5$M_{\oplus}$ more
massive hydrogen envelopes within the mass range of $\sim
7.5\times 10^{23}$--$1.5\times 10^{28}$ g can be captured from the
nebula. For studying the escape of these accumulated
hydrogen-dominated protoatmospheres, we apply a hydrodynamic upper
atmosphere model and calculate the loss rates due to the heating
by the high soft-X-ray and extreme ultraviolet (XUV) flux of the
young Sun/star.
The results of our study
indicate that under most nebula conditions `sub-Earth' and
Earth-mass planets can lose their captured hydrogen envelopes by
thermal escape during the first $\sim 100$ Myr after the disk
dissipated. However, if a nebula has a low dust depletion factor
or low accretion rates resulting in low protoplanetary
luminosities, it is possible that even protoplanets with
Earth-mass cores may keep their hydrogen envelopes during their
whole lifetime. In contrast to lower mass protoplanets, more
massive `super-Earths' that can accumulate a huge amount of nebula
gas, lose only tiny fractions of their primordial hydrogen
envelopes. Our results agree with the fact that Venus, Earth, and
Mars are not surrounded by dense hydrogen envelopes, as well as
with the recent discoveries of low density `super-Earths' that
most likely could not get rid of their dense protoatmospheres.
\end{abstract}


\section{INTRODUCTION}
During the growth from planetesimals to planetary embryos and the
formation of protoplanets in the primordial solar nebula, massive
hydrogen/helium envelopes can be captured (e.g., Mizuno et al.
1978; Wuchterl 1993;  Ikoma et al. 2000; Rafikov 2006). The capture
of nebula gas becomes important when the growing planetary embryos
grow to a stage, where gas accretion exceeds solid body accretion, so that the
massive gaseous envelopes of Jupiter-type planets can be formed
(e.g., Pollack 1985; Lissauer et al. 1995; Wuchterl 1995; 2010).

Analytical and numerical
calculations indicate that the onset of core instability occurs
when the mass of the hydrogen atmosphere around the core becomes
comparable to the core mass (e.g., Hayashi et al. 1978; Mizuno et
al. 1978; Mizuno 1980; Stevenson 1982; Wuchterl 1993; 1995; Ikoma et al.
2000; Rafikov 2006; Wuchterl 2010). This fast growth of nebula-based hydrogen
envelopes occurs if the core reaches a mass of
$\sim 10 M_{\rm \oplus}$ and a size of $\sim 2 R_{\rm \oplus}$
(e.g., Alibert et al. 2010).

As it was shown in the early pioneering studies by Mizuno et al.
(1978; 1980) and Hayashi et al. (1979) if the core mass becomes
$\geq 0.1 M_{\rm \oplus}$, depending on nebula
properties, an appreciable amount of the surrounding nebula gas
can be captured by terrestrial-type protoplanets such as
proto-Venus or proto-Earth to form an optically thick, dense
primordial hydrogen atmosphere with a mass in the order of $\sim
10^{26}$g or $\sim 653$ Earth ocean equivalent amount of hydrogen
(EO$_{\rm H}$)\footnote{1 EO$_{\rm H}\approx 1.53 \times 10^{23}$
g.}. Because Earth doesn't have such a hydrogen envelope at
present, Sekiya et al. (1980a; 1980b)
studied the evaporation of this primordial nebula-captured
terrestrial atmosphere due to irradiation of the solar EUV
radiation of the young Sun.

These authors suggested that Earth's nebula captured hydrogen
envelope of the order of $\sim 10^{26}$ g should have dissipated within a
period of $\sim 500$ Myr, because $\sim$4 Gyr ago nitrogen was
the dominant species in the atmosphere. It was found that
for removing the expected hydrogen envelope during the
time period of $\sim 500$ Myr, the EUV flux of the young Sun
should have been $\sim 200$ times higher compared to the present
day value. However, Sekiya et al. (1980a; 1980b) concluded that such a
high EUV flux during this long time period was not realistic. To overcome this problem, Sekiya et al. (1981) assumed in a follow up study
a far-UV flux, that was $\sim 100$ times higher compared to that of
the present Sun during the T-Tauri stage and its absorption occurred through photodissociation of H$_2$O molecules. Under this assumptions Earth's, protoatmosphere dissipated within $\sim 20$ Myr (Sekiya et al. 1981).

However, the
results and conclusions of these pioneering studies are based on very
rough assumptions, because decades ago, no good observational data
from disks and related nebula lifetimes, as well as astrophysical
observations of the radiation environment of very young solar-like
stars have been available. Since that time several multiwavelength
satellite observations of the radiation environment of young
solar-like stars indicate that the assumed EUV and far-UV flux
values by Sekiya et al. (1980b; 1981) have been too high and/or
lasted for shorter timescales (Guinan et al. 2003; Ribas et al.
2005; G\"{u}del 2007; Claire et al. 2012; Lammer et al. 2012; and
references therein).

The main aim of this study is to investigate the capture and
loss of nebula-based hydrogen envelopes from protoplanets
with `sub-`to `super-Earth' masses, during
the XUV saturated phase of a Sun-like star (e.g., Ayres 1997; G\"{u}del et al.
1995; G\"{u}del 1997; G\"{u}del et al. 1997; Guinan et al. 2003;
Ribas et al. 2005; Claire et al. 2012),
in the habitable zone (HZ) at 1 AU. In Sect. 2 we discuss the latest stage of
our knowledge regarding the early solar/stellar radiation
environment from the time when the nebula dissipated
and protoplanetary atmospheres have been exposed freely to the
high solar XUV radiation field. In Sect. 3 we study the formation
of nebula captured hydrogen envelopes on protoplanets with an
Earth-like mean core density and core masses of 0.1$M_{\rm \oplus}$,
0.5$M_{\rm \oplus}$, 1$M_{\rm \oplus}$, 2$M_{\rm \oplus}$,
3$M_{\rm \oplus}$, and 5$M_{\rm \oplus}$ by solving the
hydrostatic structure equations for the protoplanetary nebula for
different dust grain depletion factors, planetesimal accretion
rates, and corresponding protoplanetary luminosities. In Sect. 4
we study the heating and expansion of the captured hydrogen envelopes
by the XUV radiation, as well as the
related thermal escape rates during the XUV saturation phase of the young Sun/stars
by applying a time-dependent numerical algorithm, that solves the system of 1-D fluid
equations for mass, momentum, and energy conservation. Finally we
discuss our results and their implications for the formation of
early Venus, Earth, Mars, terrestrial exoplanets, and habitability
in general.
\section{THE EARLY XUV ENVIRONMENT OF SOLAR-LIKE STARS}
\label{XUV}
\subsection{The XUV activity during the stellar saturation phase}
The evolution of the solar high-energy radiation has been studied in detail within the `Sun in Time' program (e.g., Ribas et al. 2005; G\"udel 2007). At ages higher than about 0.1~Gyr, the activity of Sun-like stars decreases because of magnetic braking by the stellar wind. Data from X-rays to UV from a sample of solar analogs of different ages provided power laws of the solar fluxes in different wavelength bands within the XUV range (1--1200~\AA) as a function of age. Recently, Claire et al. (2012) extended this study to longer wavelengths and provided a code with calculates full spectra from X-rays to the infrared for a range of stellar ages. Using their model, we calculate solar fluxes in several wavelength bands back to an age of 0.1~Gyr, which corresponds approximately to the end of the saturation period for a Sun-like star (Fig.~1). We normalize the derived XUV fluxes to the present-day (4.56~Gyr) value of $5.27\,\mathrm{erg\,cm^{-2}\,s^{-1}}$ as calculated with the model of Claire et al. (2012), which corresponds to the Sun's emission during activity maximum, to obtain the evolution of the flux enhancement factor. This fiducial value is slightly higher than the moderate solar value of $4.64\,\mathrm{erg\,cm^{-2}\,s^{-1}}$ adopted by Ribas et al. (2005), but the evolution of the flux enhancement factor calculated with their power law is almost identical. We prefer the spectral model mainly because we are interested in the evolution of other wavelength bands for which no power laws were given by Ribas et al. (2005). Note that the XUV flux of the present-day Sun is highly variable and may change by up to an order of magnitude. In this section we adopt the fiducial values derived from the Claire et al. (2012) model for 4.56~Gyr to simply set the Sun at one in Fig.~1.

During the saturation phase of a solar-like G-type star ($\leq 0.1$~Gyr), the stellar X-ray luminosity is saturated at a level of about 0.1\% of the bolometric luminosity (Pizzolato et al. 2003; Jackson et al. 2012). Before the Sun reached the main sequence, its bolometric luminosity varied strongly according to stellar evolution models. Hence, the high-energy emission is expected to roughly follow the evolution of the bolometric luminosity. This is consistent with X-ray observations of solar-type stars in young clusters. To estimate the Sun's XUV emission during the saturation phase, we use an evolution model from Baraffe et al. (1998) for a solar-mass star and adopt an X-ray saturation level of $L_\mathrm{X,sat}=10^{(-3.2\pm0.3)}L_\mathrm{bol}$ from Pizzolato et al. (2003) for stars of about one solar mass. We assume that most of the XUV emission during this phase is emitted in X-rays, however, the unknown contribution at longer wavelengths may increase the scaling factor. We caution that the estimated enhancement factor during this period is very uncertain. However, in agreement with the normalized values shown in Fig. 1 we apply for the thermal escape model below, an average XUV enhancement factor during the saturation phase of the solar-like star that is 100 times higher compared to the value at 1 AU of today's Sun.
\begin{figure}
\includegraphics[width=0.94\columnwidth]{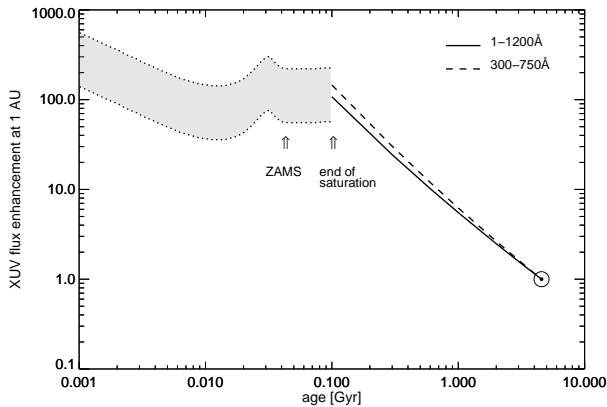}
\caption{Evolution of the solar high-energy emission in different wavelength bands calculated with the model of Claire et al. (2012). All fluxes are taken at 1~AU and normalized to the present solar values of $[5.27,1.69]\,\mathrm{erg\,cm^{-2}\,s^{-1}}$ for the XUV, EUV bands. Before 0.1~Gyr the Sun's XUV evolution is approximated by the modeled evolution of the bolometric luminosity of a star of $1M_\odot$ (Baraffe et al. 1998) and an X-ray saturation level of $10^{-3.2\pm0.3}L_\mathrm{bol}$ (Pizzolato et al. 2003), denoted by the shaded area. Arrows indicate the zero-age main sequence (ZAMS) and the approximate end of the saturation phase.}
\end{figure}

\subsection{The dissipation time of the nebula gas}
The capture of primordial atmospheres by accreting protoplanetary cores
depend mainly on the nebula lifetime, nebula parameters such as the dust-depletion factor,
core mass, and core accretion luminosity.
The life time of the nebula can be estimated from infrared observations of T-Tauri stars in
star forming regions, that are sensitive
to the presence of circumstellar material.
Such observations indicate that disks evaporate and disappear on different timescales
(e.g., Hillenbrand 2006; Montmerle et al. 2006).
Inner disks where the regions are warm enough to radiate at near-IR wavelengths
disappear after a few Myrs (Hillenbrand 2006). The fraction of stars with discs
is consistent with 100\% at an age of 1 Myr, and drops to less than 10\% after 10 Myr.
There are some observations of older disks but their masses are too
low to form a Jupiter-like gas giant. Therefore, the common view is that
giant planets must form on timescales which are $< 10$ Myr
(Montmerle et al. 2006; and references therein). Because of this fact, one can expect that
cores with masses from `sub'- to `super-Earths' can also be formed during the time before the
nebula gas evaporates.
\section{ORIGIN OF NEBULA-CAPTURED PROTOATMOSPHERES}
\label{proto}
\begin{table*}
\renewcommand{\baselinestretch}{1}
\caption{Masses and surface properties of captured protoatmospheres around rocky cores of
0.1$M_{\rm \oplus}$, 0.5$M_{\rm \oplus}$, 1$M_{\rm \oplus}$, 2$M_{\rm \oplus}$, 3$M_{\rm \oplus}$,
and 5$M_{\rm \oplus}$ using a dust depletion factor $f$ of 0.001.
The relative accretion rate $\dot{M}_{\rm acc}/M_{\rm pl}$ is a free parameter and defines a planetary luminosity $L_{\rm pl}$.
$T_{\rm s}$ and $P_{\rm s}$ are the surface temperature and pressure, the
atmospheric mass $M_{\rm atm}$ is given in g and Earth ocean equivalent amounts of hydrogen EO$_{\rm H}$.
The hydrogen envelope mass fraction is defined as $f_{\rm env} = M_{\rm atm}/M_{\rm pl}$, where $M_{\rm pl} = (M_{\rm atm}+M_{\rm core})$.}\label{atmo-mods_dust0.001}
\begin{center}
\begin{tabular}{ccccccccc}
$M_{\rm core}$ [$M_{\rm \oplus}$] & $R_{\rm pl}$ [$R_{\rm \oplus}$]  &  $\frac{\dot{M}_{\rm acc}}{M_{\rm pl}}$ [yr$^{-1}$]  &  $L_{\rm pl}$ [erg s$^{-1}$]  &  $T_{\rm s}$ [K]  &  $P_{\rm s}$ [bar]  &$M_{\rm atm}$ [g]& $M_{\rm atm}$ [EO$_{\rm H}$] & $f_{\rm env}$\\\hline\hline
0.1  &  0.46 &           $10^{-5}$       &  $2.54 \times 10^{25}$  &  961   &  0.155  &  $2.13 \times 10^{21}$  & 0.014&$3.6 \times 10^{-6}$ \\
0.5  &  0.79 &           $10^{-5}$       &  $3.71 \times 10^{26}$  &  2687  &  4.703  &  $2.19 \times 10^{23}$  & 1.429&$7.3 \times 10^{-5}$ \\
1    &  1    &           $10^{-5}$       &  $1.18 \times 10^{27}$  &  3594  &  24.81  &  $1.47 \times 10^{24}$  & 9.608&$2.5 \times 10^{-4}$ \\
2    &  1.26 &           $10^{-5}$       &  $3.74 \times 10^{27}$  &  4780  &  168.8  &  $1.01 \times 10^{25}$  & 66.01&$8.4 \times 10^{-4}$ \\
3    &  1.44 &           $10^{-5}$       &  $7.35 \times 10^{27}$  &  5810  &  558.2  &  $3.24 \times 10^{25}$  & 211.5&$1.8 \times 10^{-3}$ \\
5    &  1.71 &           $10^{-5}$       &  $1.72 \times 10^{28}$  &  7792  &  2503   &  $1.46 \times 10^{26}$  & 956.2&$4.9 \times 10^{-3}$ \\
\hline
0.1  &  0.46 &           $10^{-6}$       &  $2.54 \times 10^{24}$  &  871   &  0.768  &  $7.18 \times 10^{21}$  & 0.047&$1.2 \times 10^{-5}$ \\
0.5  &  0.79 &           $10^{-6}$       &  $3.71 \times 10^{25}$  &  2749  &  26.73  &  $9.60 \times 10^{23}$  & 6.275&$3.2 \times 10^{-4}$ \\
1    &  1    &           $10^{-6}$       &  $1.18 \times 10^{26}$  &  3802  &  134.7  &  $6.82 \times 10^{24}$  & 44.6&$1.1 \times 10^{-3}$ \\
2    &  1.26 &           $10^{-6}$       &  $3.74 \times 10^{26}$  &  5224  &  833.8  &  $4.88 \times 10^{25}$  & 319&$4.1 \times 10^{-3}$ \\
3    &  1.44 &           $10^{-6}$       &  $7.35 \times 10^{26}$  &  6423  &  2561   &  $1.58 \times 10^{26}$  & 1030&$8.7 \times 10^{-3}$ \\
5    &  1.71 &           $10^{-6}$       &  $1.72 \times 10^{27}$  &  8690  &  10377  &  $7.12 \times 10^{26}$  & 4653&0.023               \\
\hline
0.1  &  0.46 &           $10^{-7}$       &  $2.54 \times 10^{23}$  &  770   &  3.329  &  $2.29 \times 10^{22}$  & 0.15&$3.8 \times 10^{-5}$ \\
0.5  &  0.79 &           $10^{-7}$       &  $3.71 \times 10^{24}$  &  2729  &  154.4  &  $4.32 \times 10^{24}$  & 28.2&$1.4 \times 10^{-3}$ \\
1    &  1    &           $10^{-7}$       &  $1.18 \times 10^{25}$  &  3972  &  744.6  &  $3.21 \times 10^{25}$  & 210&$5.4 \times 10^{-3}$ \\
2    &  1.26 &           $10^{-7}$       &  $3.74 \times 10^{25}$  &  5601  &  4126   &  $2.40 \times 10^{26}$  & 1565&  0.020               \\
3    &  1.44 &           $10^{-7}$       &  $7.35 \times 10^{25}$  &  6955  &  11742  &  $8.06 \times 10^{26}$  & 5270& 0.043               \\
5    &  1.71 &           $10^{-7}$       &  $1.72 \times 10^{26}$  &  9494  &  45193  &  $4.38 \times 10^{27}$  & 28620& 0.13                \\
\hline
0.1  &  0.46 &           $10^{-8}$       &  $2.55 \times 10^{22}$  &  677   &  10.70  &  $5.99 \times 10^{22}$  & 0.392& $1.0 \times 10^{-4}$ \\
0.5  &  0.79 &           $10^{-8}$       &  $3.72 \times 10^{23}$  &  2648  &  772.9  &  $1.79 \times 10^{25}$  & 116.7& $5.9 \times 10^{-3}$ \\
1    &  1    &           $10^{-8}$       &  $1.18 \times 10^{24}$  &  3987  &  3929   &  $1.53 \times 10^{26}$  & 1002&  0.025               \\
2    &  1.26 &           $10^{-8}$       &  $3.74 \times 10^{24}$  &  5722  &  21892  &  $1.52 \times 10^{27}$  & 9934&  0.11                \\
3    &  1.44 & $1.41 \times 10^{-8}\,^a$ &  $1.04 \times 10^{25}$  &  7291  &  63099  &  $8.67 \times 10^{27}$  & 56640&  0.33                \\
5    &  1.71 & $5.11 \times 10^{-8}\,^a$ &  $8.81 \times 10^{25}$  &  9955  &  96262  &  $1.62 \times 10^{28}$  & 105816&  0.35                \\
\hline
\end{tabular}
\end{center}
$^a)$ Accretion rate increased to obtain a stationary solution.
\normalsize
\end{table*}
\begin{table*}
\renewcommand{\baselinestretch}{1}
\caption{Same as Tab.~\ref{atmo-mods_dust0.001} for a dust depletion factor $f$ of 0.01}.
\label{atmo-mods_dust0.01}
\begin{center}
\begin{tabular}{ccccccccc}
$M_{\rm core}$ [$M_{\rm \oplus}$]  & $R_{\rm pl}$ [$R_{\rm \oplus}$]  & $\frac{\dot{M}_{\rm acc}}{M_{\rm pl}}$  [yr$^{-1}$] &  $L_{\rm pl}$ [erg s$^{-1}$]  &  $T_{\rm s}$ [K]  &  $P_{\rm s}$ [bar]  &  $M_{\rm atm}$ [g]  & $M_{\rm atm}$ [EO$_{\rm H}$] & $f_{\rm env}$\\
\hline
\hline
0.1  &  0.46 &           $10^{-5}$       &  $2.54 \times 10^{25}$  &  924   & 0.033  &  $3.41 \times 10^{20}$  &0.0022  &$5.7 \times 10^{-7}$ \\
0.5  &  0.79 &           $10^{-5}$       &  $3.71 \times 10^{26}$  &  2412  & 1.299  &  $3.47 \times 10^{22}$  &0.227  &$1.2 \times 10^{-5}$ \\
1    &  1    &           $10^{-5}$       &  $1.18 \times 10^{27}$  &  3307  & 8.395  &  $2.57 \times 10^{23}$  &1.682  &$4.3 \times 10^{-5}$ \\
2    &  1.26 &           $10^{-5}$       &  $3.74 \times 10^{27}$  &  4436  & 68.93  &  $2.12 \times 10^{24}$  &13.84  &$1.8 \times 10^{-4}$ \\
3    &  1.44 &           $10^{-5}$       &  $7.35 \times 10^{27}$  &  5399  & 254.0  &  $8.09 \times 10^{24}$  &52.86  &$4.5 \times 10^{-4}$ \\
5    &  1.71 &           $10^{-5}$       &  $1.72 \times 10^{28}$  &  7210  & 1288   &  $4.57 \times 10^{25}$  &299  &$1.5 \times 10^{-3}$ \\
\hline
0.1  &  0.46 &           $10^{-6}$       &  $2.54 \times 10^{24}$  &  848   &  0.184  &  $2.26 \times 10^{21}$  &0.014  &$3.8 \times 10^{-6}$ \\
0.5  &  0.79 &           $10^{-6}$       &  $3.71 \times 10^{25}$  &  2516  &  8.122  &  $2.72 \times 10^{23}$  &1.776  &$9.1 \times 10^{-5}$ \\
1    &  1    &           $10^{-6}$       &  $1.18 \times 10^{26}$  &  3563  &  46.92  &  $1.96 \times 10^{24}$  &12.8  &$3.3 \times 10^{-4}$ \\
2    &  1.26 &           $10^{-6}$       &  $3.74 \times 10^{26}$  &  4865  &  334.1  &  $1.47 \times 10^{25}$  &96.14  &$1.2 \times 10^{-3}$ \\
3    &  1.44 &           $10^{-6}$       &  $7.35 \times 10^{26}$  &  5979  &  1120   &  $4.99 \times 10^{25}$  &326.2  &$2.8 \times 10^{-3}$ \\
5    &  1.71 &           $10^{-6}$       &  $1.72 \times 10^{27}$  &  8091  &  5040   &  $2.45 \times 10^{26}$  &1600  &$8.1 \times 10^{-3}$ \\
\hline
0.1  &  0.46 &           $10^{-7}$       &  $2.54 \times 10^{23}$  &  790   &  0.956  &  $8.06 \times 10^{21}$  &0.052  &$1.3 \times 10^{-5}$ \\
0.5  &  0.79 &           $10^{-7}$       &  $3.71 \times 10^{24}$  &  2590  &  49.28  &  $1.33 \times 10^{24}$  &8.673  &$4.4 \times 10^{-4}$ \\
1    &  1    &           $10^{-7}$       &  $1.18 \times 10^{25}$  &  3772  &  264.3  &  $1.01 \times 10^{25}$  &65.75  &$1.7 \times 10^{-3}$ \\
2    &  1.26 &           $10^{-7}$       &  $3.74 \times 10^{25}$  &  5304  &  1644   &  $7.66 \times 10^{25}$  &500.3  &$6.4 \times 10^{-3}$ \\
3    &  1.44 &           $10^{-7}$       &  $7.35 \times 10^{25}$  &  6577  &  4997   &  $2.57 \times 10^{26}$  &1680  & 0.014               \\
5    &  1.71 &           $10^{-7}$       &  $1.72 \times 10^{26}$  &  8956  &  19911  &  $1.24 \times 10^{27}$  &8117  &0.040               \\
\hline
0.1  &  0.46 &           $10^{-8}$       &  $2.55 \times 10^{22}$  &  717   &  4.441  &  $2.80 \times 10^{22}$  &0.183  &$4.7 \times 10^{-5}$ \\
0.5  &  0.79 &           $10^{-8}$       &  $3.72 \times 10^{23}$  &  2623  &  276.9  &  $6.42 \times 10^{24}$  &41.95  &$2.1 \times 10^{-3}$ \\
1    &  1    &           $10^{-8}$       &  $1.18 \times 10^{24}$  &  3933  &  1408   &  $5.09 \times 10^{25}$  &332.6  &$8.4 \times 10^{-3}$ \\
2    &  1.26 &           $10^{-8}$       &  $3.74 \times 10^{24}$  &  5626  &  7910   &  $4.09 \times 10^{26}$  &2670  & 0.033               \\
3    &  1.44 &           $10^{-8}$       &  $7.35 \times 10^{24}$  &  7013  &  22656  &  $1.50 \times 10^{27}$  &9797  &0.077               \\
5    &  1.71 & $1.11 \times 10^{-8}\,^a$ &  $1.91 \times 10^{25}$  &  9894  &  105653 &  $1.59 \times 10^{28}$  &103725 &0.35                \\
\hline
\end{tabular}
\end{center}
$^a)$ Accretion rate increased to obtain a stationary solution.
\normalsize
\end{table*}
\begin{table*}
\renewcommand{\baselinestretch}{1}
\caption{Same as Tab.~\ref{atmo-mods_dust0.001} for a dust depletion factor $f$ of 0.1.}
\label{atmo-mods_dust0.1}
\begin{center}
\begin{tabular}{ccccccccc}
$M_{\rm core}$ [$M_{\rm \oplus}$]  &  $R_{\rm pl}$ [$R_{\rm \oplus}$]  &  $\frac{\dot{M}_{\rm acc}}{M_{\rm pl}}$  [yr$^{-1}$] &  $L_{\rm pl}$ [erg s$^{-1}$]  &  $T_{\rm s}$ [K]  &  $P_{\rm s}$ [bar]  &  $M_{\rm atm}$ [g]  &  $M_{\rm atm}$ [EO$_{\rm H}$] & $f_{\rm env}$\\
\hline
\hline
0.1  &  0.46 &            $10^{-5}$      &  $2.54 \times 10^{25}$  &  1031  &  0.006  &  $2.65 \times 10^{19}$  &0.00017 & $4.4 \times 10^{-8}$ \\
0.5  &  0.79 &            $10^{-5}$      &  $3.71 \times 10^{26}$  &  1810  &  0.291  &  $4.48 \times 10^{21}$  &0.029 & $1.5 \times 10^{-6}$ \\
1    &  1    &            $10^{-5}$      &  $1.18 \times 10^{27}$  &  2956  &  3.287  &  $4.84 \times 10^{22}$  &0.316 &$8.1 \times 10^{-6}$ \\
2    &  1.26 &            $10^{-5}$      &  $3.74 \times 10^{27}$  &  4131  &  37.43  &  $7.26 \times 10^{23}$  &4.747 &$6.1 \times 10^{-5}$ \\
3    &  1.44 &            $10^{-5}$      &  $7.35 \times 10^{27}$  &  5081  &  160.5  &  $3.58 \times 10^{24}$  &23.37 &$2.0 \times 10^{-4}$ \\
5    &  1.71 &            $10^{-5}$      &  $1.72 \times 10^{28}$  &  6872  &  946.9  &  $2.60 \times 10^{25}$  &169.6 &$8.7 \times 10^{-4}$ \\
\hline
0.1  &  0.46 &            $10^{-6}$      &  $2.54 \times 10^{24}$  &  901   &  0.034  &  $3.53 \times 10^{20}$  &0.0023  &$5.9 \times 10^{-7}$ \\
0.5  &  0.79 &            $10^{-6}$      &  $3.71 \times 10^{25}$  &  1885  &  1.723  &  $3.77 \times 10^{22}$  &0.0246  &$1.3 \times 10^{-5}$ \\
1    &  1    &            $10^{-6}$      &  $1.18 \times 10^{26}$  &  3092  &  16.79  &  $3.54 \times 10^{23}$  &2.313  &$5.9 \times 10^{-5}$ \\
2    &  1.26 &            $10^{-6}$      &  $3.74 \times 10^{26}$  &  4440  &  165.4  &  $3.82 \times 10^{24}$  &24.95  &$3.2 \times 10^{-4}$ \\
3    &  1.44 &            $10^{-6}$      &  $7.35 \times 10^{26}$  &  5524  &  645.8  &  $1.65 \times 10^{25}$  &107.8  &$9.2 \times 10^{-4}$ \\
5    &  1.71 &            $10^{-6}$      &  $1.72 \times 10^{27}$  &  7566  &  3427   &  $1.06 \times 10^{26}$  &690.8  &$3.5 \times 10^{-3}$ \\
\hline
0.1  &  0.46 &            $10^{-7}$      &  $2.54 \times 10^{23}$  &  803   &  0.192  &  $2.29 \times 10^{21}$  &0.014  &$3.8 \times 10^{-6}$ \\
0.5  &  0.79 &            $10^{-7}$      &  $3.71 \times 10^{24}$  &  2001  &  11.36  &  $3.02 \times 10^{23}$  &1.971  &$1.0 \times 10^{-4}$ \\
1    &  1    &            $10^{-7}$      &  $1.18 \times 10^{25}$  &  3302  &  94.54  &  $2.57 \times 10^{24}$  &16.8  &$4.3 \times 10^{-4}$ \\
2    &  1.26 &            $10^{-7}$      &  $3.74 \times 10^{25}$  &  4832  &  782.6  &  $2.35 \times 10^{25}$  &153.2  &$2.0 \times 10^{-3}$ \\
3    &  1.44 &            $10^{-7}$      &  $7.35 \times 10^{25}$  &  6071  &  2737   &  $8.94 \times 10^{25}$  &584.1  &$5.0 \times 10^{-3}$ \\
5    &  1.71 &            $10^{-7}$      &  $1.72 \times 10^{26}$  &  8341  &  12608  &  $4.97 \times 10^{26}$  &3250  & 0.016               \\
\hline
0.1  &  0.46 &            $10^{-8}$      &  $2.55 \times 10^{22}$  &  728   &  1.032  &  $8.34 \times 10^{21}$  &0.054  &$1.4 \times 10^{-5}$ \\
0.5  &  0.79 &            $10^{-8}$      &  $3.72 \times 10^{23}$  &  2180  &  76.42  &  $1.65 \times 10^{24}$  &10.75 &$5.5 \times 10^{-4}$ \\
1    &  1    &            $10^{-8}$      &  $1.18 \times 10^{24}$  &  3537  &  528.3  &  $1.43 \times 10^{25}$  &93.2  &$2.4 \times 10^{-3}$ \\
2    &  1.26 &            $10^{-8}$      &  $3.74 \times 10^{24}$  &  5212  &  3736   &  $1.26 \times 10^{26}$  &826.1  & 0.010               \\
3    &  1.44 &            $10^{-8}$      &  $7.35 \times 10^{24}$  &  6542  &  11790  &  $4.69 \times 10^{26}$  &3066  & 0.026               \\
5    &  1.71 &            $10^{-8}$      &  $1.72 \times 10^{25}$  &  8974  &  48871  &  $2.73 \times 10^{27}$  &17810  & 0.084               \\
\hline
\end{tabular}
\end{center}
\normalsize
\end{table*}
Once protoplanets orbiting a host star embedded in a
protoplanetary nebula gain enough mass for their gravitational
potential to become significant relatively to the local gas
internal and turbulent motion energy, they will start to
accumulate nebula gas into their atmospheres. Hydrostatic and
spherically symmetric models of such primary hydrogen atmospheres
have been applied by \citet{Hayashi1979} and \citet{Nakazawa1985}
assuming a continuous influx of planetesimals which generated
planetary luminosity. The planetesimals are assumed to travel
through the atmosphere without energy losses so that the gained
gravitational energy has been released on the planetary surface.
More recently, \citet{Ikoma2006} improved these earlier results by
adopting a more realistic equation of state and opacities, as well
as an extended parameter space. In particular, this is the first
study which uses detailed dust opacity data, i.e.\ includes
evaporation/condensation of various dust species which directly
affects the atmospheric structure.

To estimate the amount of nebula gas collected by a protoplanet
from the surrounding protoplanetary disk, we considered a series
of solid (rocky) cores with masses of 0.1, 0.5, 1, 2, 3, and
5~$M_{\rm \oplus}$ and an average core density $\rho_{\rm core}$
of 5.5 g cm$^{-3}$ and computed self-gravitating, hydrostatic,
spherically symmetric atmospheres around those bodies.
Spherical symmetry is a good assumption in
the inner parts of the planetary atmosphere, but less so in the
outer part, where the atmospheric structure melds into the
circumstellar disk. Therefore we fixed the outer boundary of our
models at the point where the gravitational domination of the
planet will finally break down, i.e.\ the Hill radius defined as
\begin{equation}
R_{\rm H} = \left( \frac{M_{\rm pl} }{3 M_{\star}}\right)^{1/3} d \;,
\end{equation}
where $d$ is the semi-major axis of the planetary orbit. Because
we are interested in studying the capture and losses of hydrogen
envelopes within the HZ of a Sun-like star, we assume
$d$=1 AU and $M_{\star} = M_{\odot}$ for all model runs.
When evaluating the captured atmospheric mass, we included only the
atmospheric structure up to the Bondi radius
\begin{equation}
R_{\rm B} = \frac{G\;M_{\rm pl} }{h} \;.
\end{equation}
(which usually is about an order of magnitude smaller than $R_{\rm H}$).
Beyond $R_{\rm B}$
the enthalpy $h$ of the gas exceeds the
gravitational potential, thus the gas is not effectively bound to
the planet and will escape once the nebula disappears.

For the physical conditions in the circumstellar disk (i.e.\ at
$R_{\rm H}$), in particular temperature and density, we assume
T~=~200~K and $\rho = 5 \times 10^{-10}$ g cm$^{-3}$, in
accordance with models of the initial mass planetary nebula
\citep{Hayashi1981}. We would like to stress the point that in our
models the atmospheric masses of substantial atmospheres (with the
surface pressure greater than $\sim 1$ bar) are rather insensitive
to the location of the outer boundary (i.e. $M_{\star}$ and $d$)
and to the physical conditions at this location. This fact
is in agreement with the earlier studies of planetary atmospheres
with radiative outer envelopes \citep[e.g.][]{Stevenson1982,
Wuchterl1993, Rafikov2006}.

The temperature of the planetary atmosphere is determined by the
planetary luminosity, which in turn is connected to the continuous
accretion of planetesimals onto the planet. The observed lifetimes
of protoplanetary disks of some Myrs give some constrains on the
formation times of the planetary cores, accordingly we adopted
mass accretion rates $\dot{M}_{\rm acc}/M_{\rm pl}$ of
$10^{-5}{\rm yr}^{-1}$, $10^{-6}{\rm yr}^{-1}$, and $10^{-7}{\rm
yr}^{-1}$. We assume that the energy release due to accretion
occurs exclusively on the planetary surface, yielding a luminosity
at the bottom of the atmosphere of
\begin{equation}
L_{\rm pl} \simeq G M_{\rm pl} \dot{M}_{\rm acc} \left( \frac{1}{r_{\rm pl}} - \frac{1}{r_{\rm H}}\right) \;.
\end{equation}
Apart from the luminosity generated by accretion, it is also
reasonable to assume that the planetary core stores a fraction of
the internal energy generated in the formation process
\citep{Elkins-Tanton2008, Hamano2013} which is a quite recent
event, in any case for a planet in the nebula phase. Therefore,
the cooling of the core may generate a heat flux, depending on
energy transport physics in the interior, contributing to the
luminosity on the planetary surface. Another contribution to the
planetary luminosity comes from radiogenic heat production which
is assumed to be proportional to the core mass by using the
reference value of $10^{21}$ erg s$^{-1}$ $M_{\rm \oplus}^{-1}$
for the young Earth \citep{Stacey1992}.

In order to obtain a measure for the uncertainties involved in our
modelling, we investigated a parameter space spanned by the
planetary luminosity $L_{\rm pl}$ and the dust depletion factor
$f$. The dust depletion factor $f$ describes the abundance of dust
relative to the amount of dust that would condensate from the gas in
equilibrium conditions.
In a hydrostatic structure, large dust grains will decouple dynamically
from the gas and may
thus fall (rain) toward the surface.
This effect is crudely modelled by adjusting the dust depletion factor $f$.
It seems clear that in general $f$ should be $<1$, but apart from that,
the magnitude of $f$ remains highly uncertain.
As dust opacities dominate over gas opacities usually by many
orders of magnitude, $f$ in effect scales the overall opaqueness
of the atmosphere. We selected these two parameters as they have a
very significant effect on the atmospheric structure, and yet are
very poorly constrained both from theory and observations.

The hydrostatic structure equations were integrated by using the
TAPIR-Code (short for {\it T}he {\it a}da{\it p}tive, {\it
i}mplicit {\it R}HD-Code). Convective energy transport is included
in TAPIR by an equation for the turbulent convective energy
following concepts of \citet{Kuhfuss1987}.
For the gas and dust opacity we used tables
from \citet{Freedman2008}, and \citet{Semenov2003}, respectively;
for the equation of state we adopted the results of
\citet{Saumon1995}. Both, for the gas opacity and the equation of
state we assumed solar composition and ignored chemical
inhomogeneities.

Tables \ref{atmo-mods_dust0.001} to \ref{atmo-mods_dust0.1}
summarize the constitutive parameters and integral properties of
the computed atmosphere models. Ignoring more subtle effects
caused by dust physics, opacities and convective transport, the
general trends in the results for the envelope masses can be
understood from the fact that planetary atmospheres are supported
from collapse by both the density gradient and the temperature
gradient. The amount of thermal support of a planetary atmosphere
depends, (1) on the planetary luminosity, i.e.\ the accretion
rate, and, (2) on the dust depletion factors $f$ defining the
opacity and thus the insulation properties of the envelope. For a
fixed atmospheric structure, more dust increases the opaqueness of
the atmosphere which thus becomes warmer. Accordingly, envelopes
with large dust depletion factors $f$ are more thermally supported
and less massive.

Obviously, increasing the accretion rate also leads to more
thermal support and smaller atmospheric masses. Inspection of
Tables \ref{atmo-mods_dust0.001} to \ref{atmo-mods_dust0.1} shows
that Earth-mass rocky cores capture hydrogen envelopes between
$\sim 5\times 10^{22}$ and $\sim 1.5\times 10^{24}$ g (equivalent
to $\sim 0.3$--10 EO$_{\rm H}$) for a high accretion rate of
$\dot{M}_{\rm acc}/M_{\rm pl} = 10^{-5}$ yr$^{-1}$, whereas a low
accretion rate of $\dot{M}_{\rm acc}/M_{\rm pl} = 10^{-8}$ yr$^{-1}$
leads to much more massive envelopes between $\sim 1.5\times
10^{25}$ and $\sim 1.5 \times 10^{26}$ g ($\sim 100$--1000 EO$_{\rm
H}$).

The most massive hydrogen envelopes are accumulated around rocky
cores within the so-called `super-Earth' domain for low accretion
rates and small dust depletion factors and can reach up to several
$\sim 10^3$ to $\sim 10^4$ EO$_{\rm H}$. In some cases of more
massive planetary cores we also encountered the upper mass limit
of stationary atmospheres, i.e.\ the critical core mass. For these
model runs we increased the accretion rate (as indicated in the
tables), and thus the luminosity, to stay just above the critical
limit. Such envelopes have hydrogen masses equivalent from
$\sim 5 \times$10$^4$ to $\sim 10^5$ EO$_{\rm H}$. Any further
reduction of the luminosity of these planets would push them into
the instability regime, causing a collapse of an atmosphere and
nebula gas and putting them onto the formation path of a gas giant
planet.

Our results, as presented in Tables \ref{atmo-mods_dust0.001} to
\ref{atmo-mods_dust0.1}, are in reasonable agreement with those of
\citet{Ikoma2006} considering the differences in input physics, in
particular, dust opacity and in the implementation of the outer
boundary condition. \citet{Ikoma2006} placed the outer boundary of
their models at the Bondi radius (in fact at the minimum of
$r_{\rm H}$ and $r_{\rm B}$, but this turns out to be equal to the
latter in all relevant cases) and accordingly implemented the
physical outer boundary conditions (nebula density and
temperature) much closer to the planet than in our models. Even
though, as noted above, the outer boundary conditions only
moderately affect the solution, this leads to somewhat more
massive atmospheres.
\begin{figure}
\includegraphics[width=0.94\columnwidth]{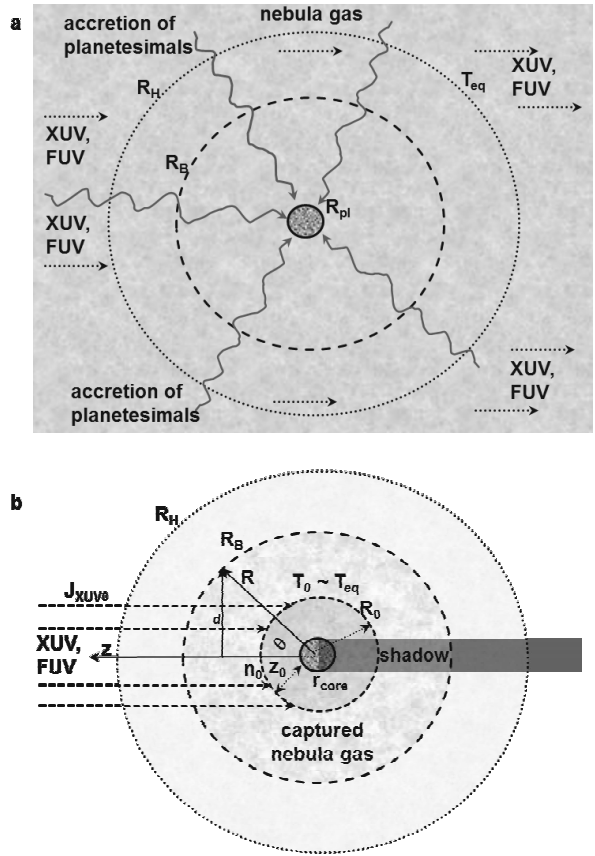}
\caption{Illustration of the protoplanet inside the nebula gas and
after its dissipation. Panel a: shaded area gas in the XUV and FUV
exposed evaporating planetary disk; corrugated arrows symbolize
the influx of accreting material; $T_{\rm eq}$ is the equilibrium
temperature of the nebula in the planet's orbital location;
$R_{\rm H}$, $R_{\rm B}$, and $R_{\rm pl}$ correspond to the Hill
radius, the Bondi radius and the planetary core radius,
respectively. Panel b: illustrates the contracted protoatmosphere
after the nebula is dissipated and the influx of planetesimals
decreased, with radius $R_0$, number density $n_0$, and
temperature $T_0$ that resembles the equilibrium
temperature $T_{\rm eq}$ at the mesopause or the
base of the thermosphere.}
\end{figure}

Finally, one should also note that the protoatmosphere models
described in this section rely on the assumption that the planetary envelopes
are in a stationary condition, so that the dynamical and thermal time scale
of the envelope are sufficiently separated both from the cooling time scale
of the solid core and the evolution time scale of the protoplanetary disk.
This assumption is not necessarily fulfilled in real planetary atmospheres.
To improve on the stationary models, further numerical development are planed
in the future might allow for the time-dependent, dynamic simulation of
such protoatmospheres (St\"{o}kl et al. 2014)
and their interaction with the solid core and the disk-nebula
environment.

\section{XUV HEATING AND ESCAPE OF CAPTURED HYDROGEN ENVELOPES}
\label{escape} As long as the protoplanet with its attracted
hydrogen envelope is embedded in the surrounding nebula gas, the
envelope is not strongly affected by evaporation. Fig. 2a
illustrates this scenario. The protoplanet is surrounded by the
dissipating nebula gas, which is also attracted by the gravity of
the rocky core $R_{\rm pl}$ up to the Hill radius $R_{\rm H}$. At
this location the gas temperature resembles more or less the
equilibrium temperature $T_{\rm eq}$ of the nebula gas that is
related to the orbital location of the planet and the luminosity
of the host star. As discussed in Sect. 3, during this early stage
the protoplanet collides with many planetesimals and planetary
embryos so that its surface is very hot. When the nebula is blown
away by the high XUV fluxes during the
T-Tauri phase of the young star, most of the nebula gas which is
captured within the Bondi sphere $R_{\rm B}$ remains and can be
considered as the protoatmosphere or captured hydrogen envelope
studied in this work.

After the dissipation phase of the gas in the disk, and the
release of energy of accreting planetesimals at the bottom of the
atmosphere decreased, the protoatmosphere contracts (e.g., Sekiya et al. 1980b; Mordasini
et al. 2012; Batygin \& Stevenson 2013) and results in
a smaller radius $R_0$ with  $T_{\rm eq} \sim T_0 \sim T_{\rm eff}$ (Marcq 2012).
As illustrated in Fig. 2b, this phase
marks the time when a protoplanet is released from the surrounding
nebula, so that its captured hydrogen envelope is exposed to the
XUV radiation of the young star and the thermal escape starts.

In the following escape studies of the captured hydrogen envelopes,
we assume that the accretion on the core of the protoplanet mainly finished,
so that the body can be considered under thermal equilibrium and parameters
such as core temperature, radiation of the host star remain constant.
\subsection{Hydrodynamic upper atmosphere model: boundary conditions and simulation domain}
Because hydrogen is a light gas and the protoplanet is exposed to an XUV flux which is $\sim$100 times higher compared to today's solar value,
the upper atmosphere will not be in hydrostatic equilibrium but will hydrodynamically expand up to several planetary radii
(e.g., Watson et al. 1981; Tian et al. 2005; Erkaev et al. 2013a; Lammer et al. 2013a).

To study the XUV-heated upper atmosphere structure and thermal
escape rates of the hydrogen envelopes from protoplanets with
masses of 0.1--5$M_{\rm \oplus}$ we apply an energy absorption and
1-D hydrodynamic upper atmosphere model, that is described in
detail in Erkaev et al. (2013a) and Lammer et al. (2013a). The
model solves the system of the hydrodynamic equations for mass,
\begin{equation}
\frac{\partial \rho R^2}{\partial t} + \frac{\partial \rho v R^2}{\partial R}= 0,
\end{equation}
momentum,
\begin{eqnarray}
\frac{\partial \rho v R^2}{\partial t} + \frac{\partial \left[ R^2 (\rho v^2+P)\right]}{\partial R} =\rho g R^2 + 2P R,
\end{eqnarray}
and energy conservation
\begin{eqnarray}
\frac{\partial R^2\left[\frac{\rho v^2}{2}+\frac{P}{(\gamma-1)}\right]}{\partial t}
+\frac{\partial v R^2\left[\frac{\rho v^2}{2}+\frac{\gamma P}{(\gamma - 1)}\right]}{\partial R}=\nonumber\\
\rho v R^2 g + q_{\rm XUV} R^2.
\end{eqnarray}
The distance $R$ corresponds to the radial distance from the
centre of the protoplanetary core, $\rho, P, T, v$ are the mass
density,pressure, temperature and velocity of the nonhydrostatic
outward flowing bulk atmosphere. $\gamma$ is the polytropic index,
$g$ the gravitational acceleration and $q_{\rm XUV}$ is the XUV
volume heating rate.
\begin{figure*}
\begin{center}
\includegraphics[width=0.9\columnwidth]{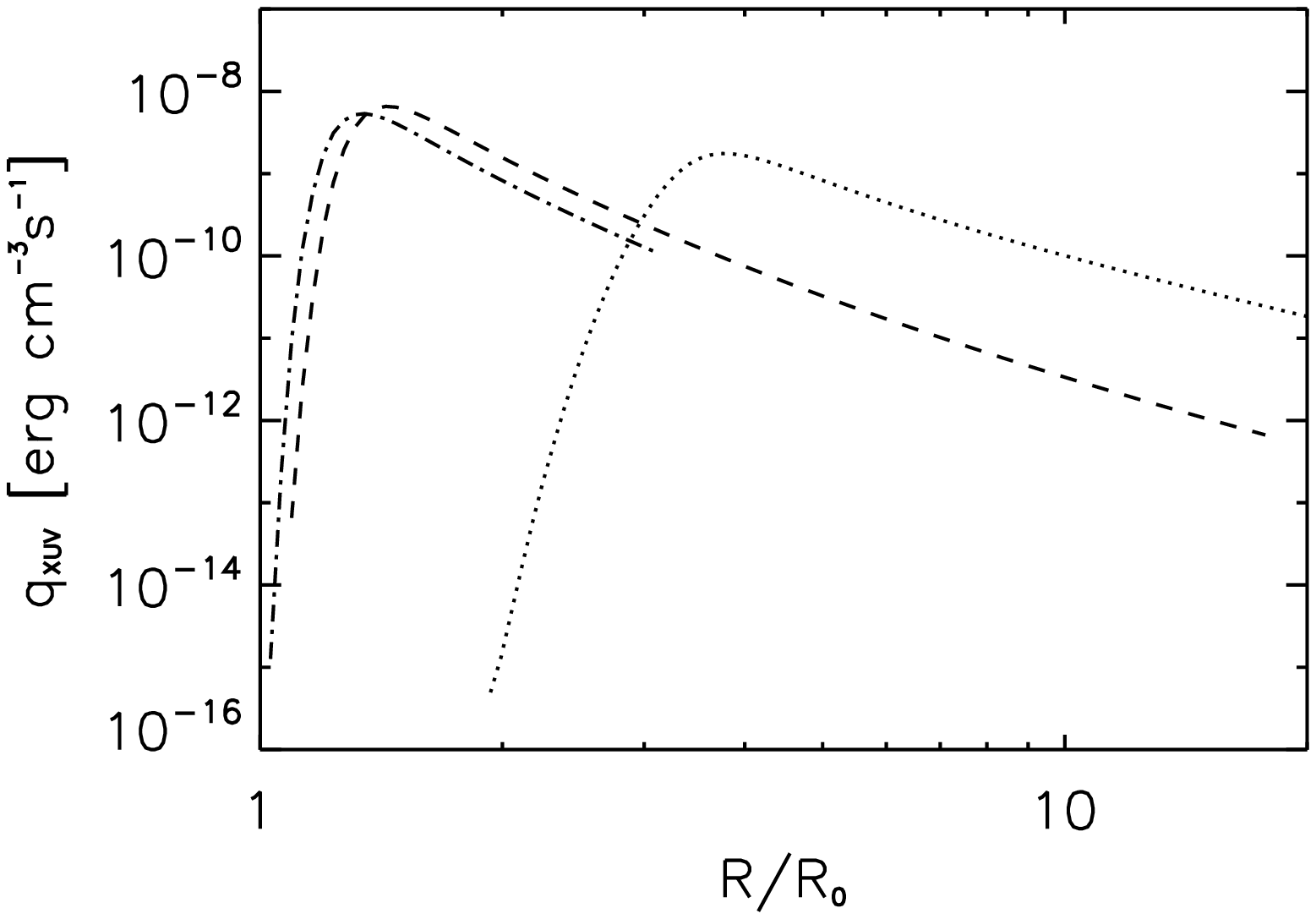}
\includegraphics[width=0.9\columnwidth]{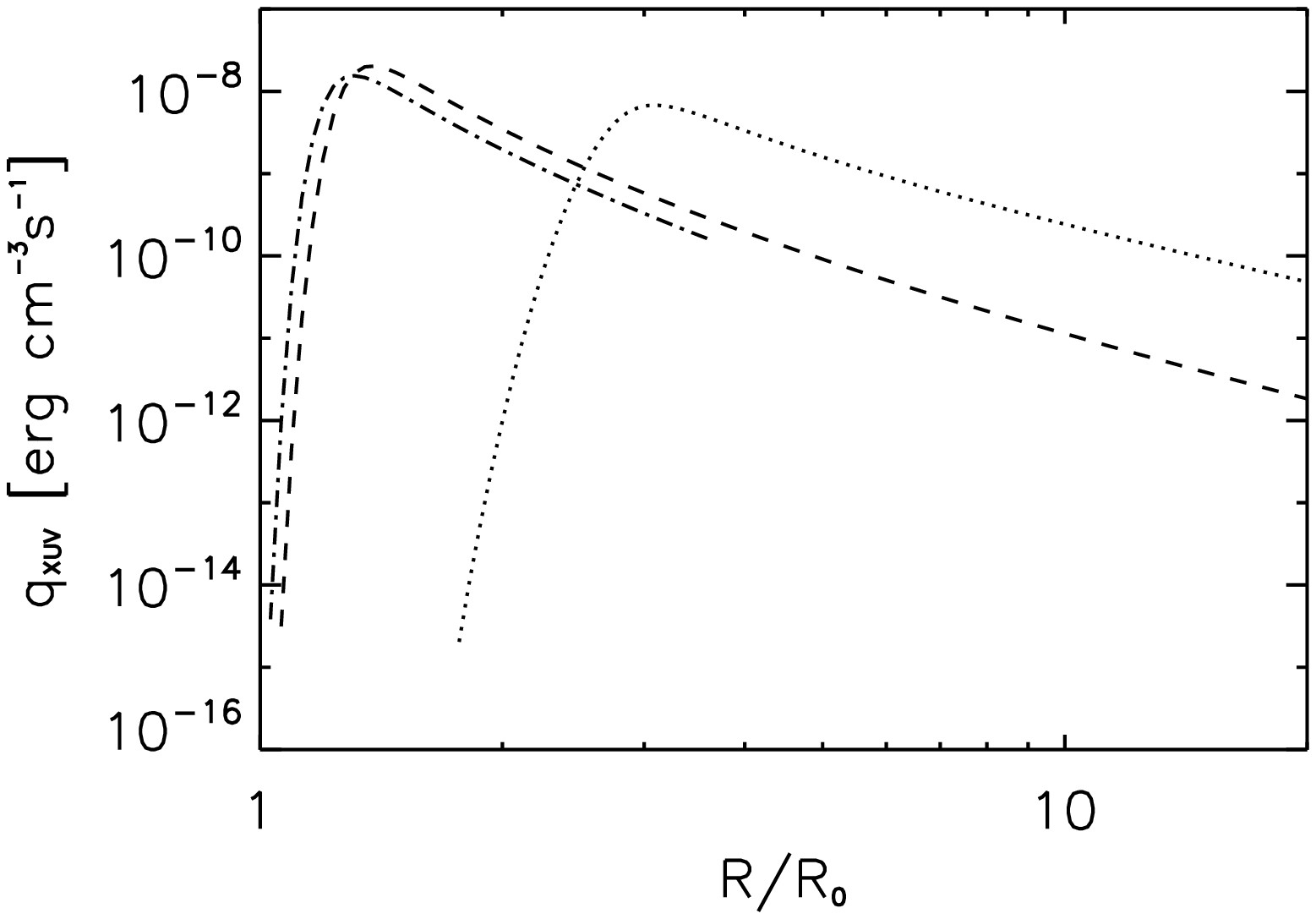}
\includegraphics[width=0.9\columnwidth]{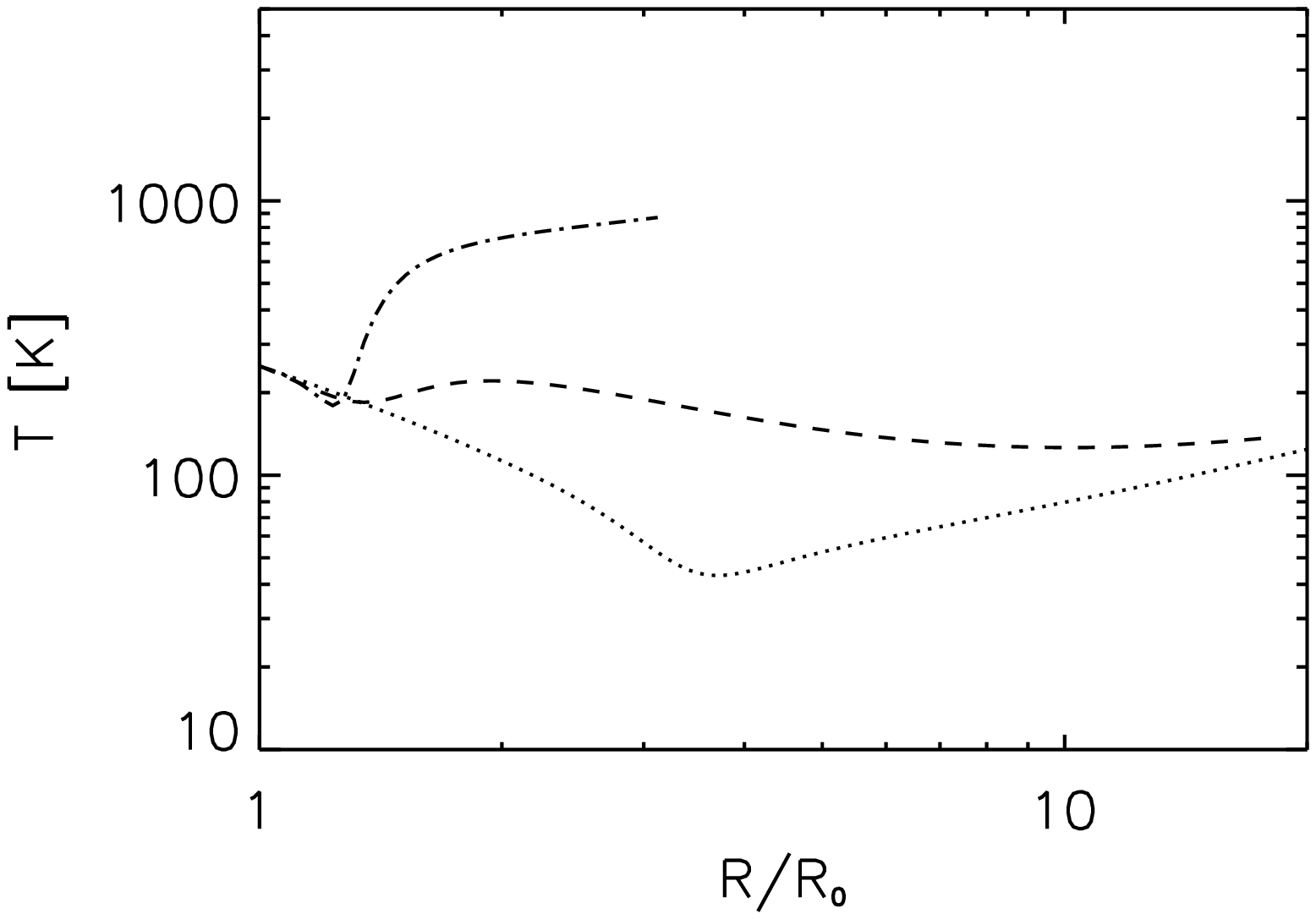}
\includegraphics[width=0.9\columnwidth]{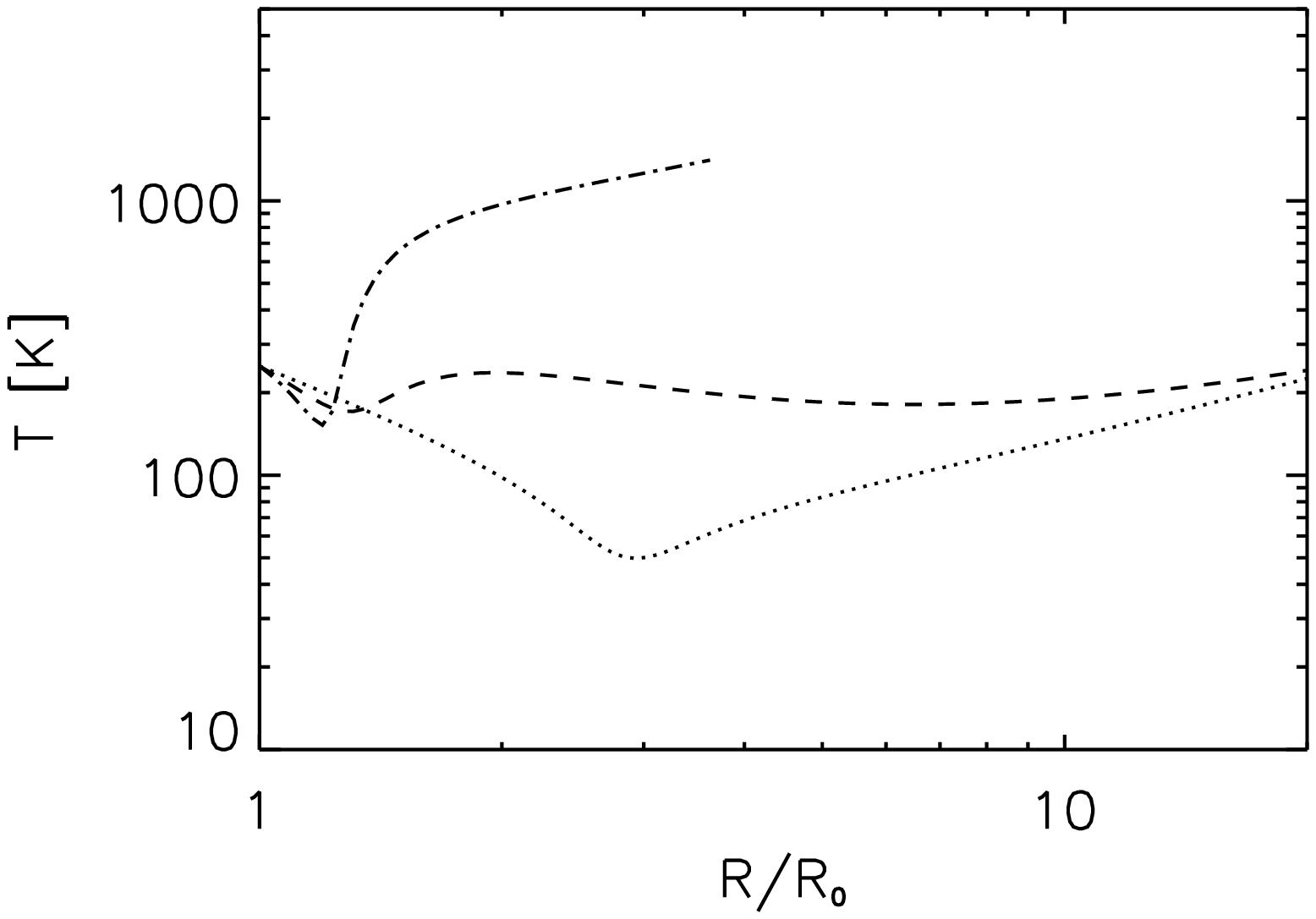}
\includegraphics[width=0.9\columnwidth]{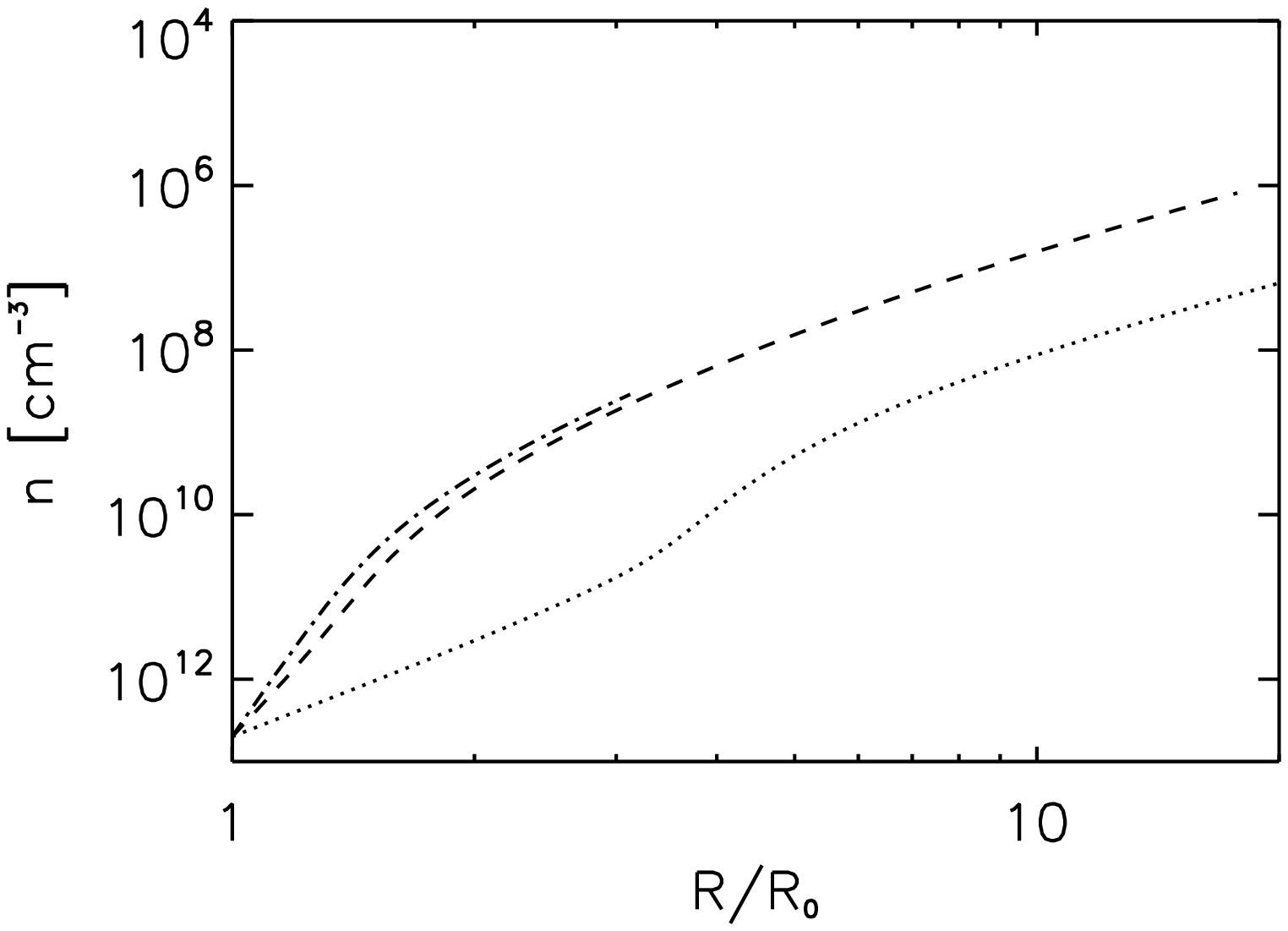}
\includegraphics[width=0.9\columnwidth]{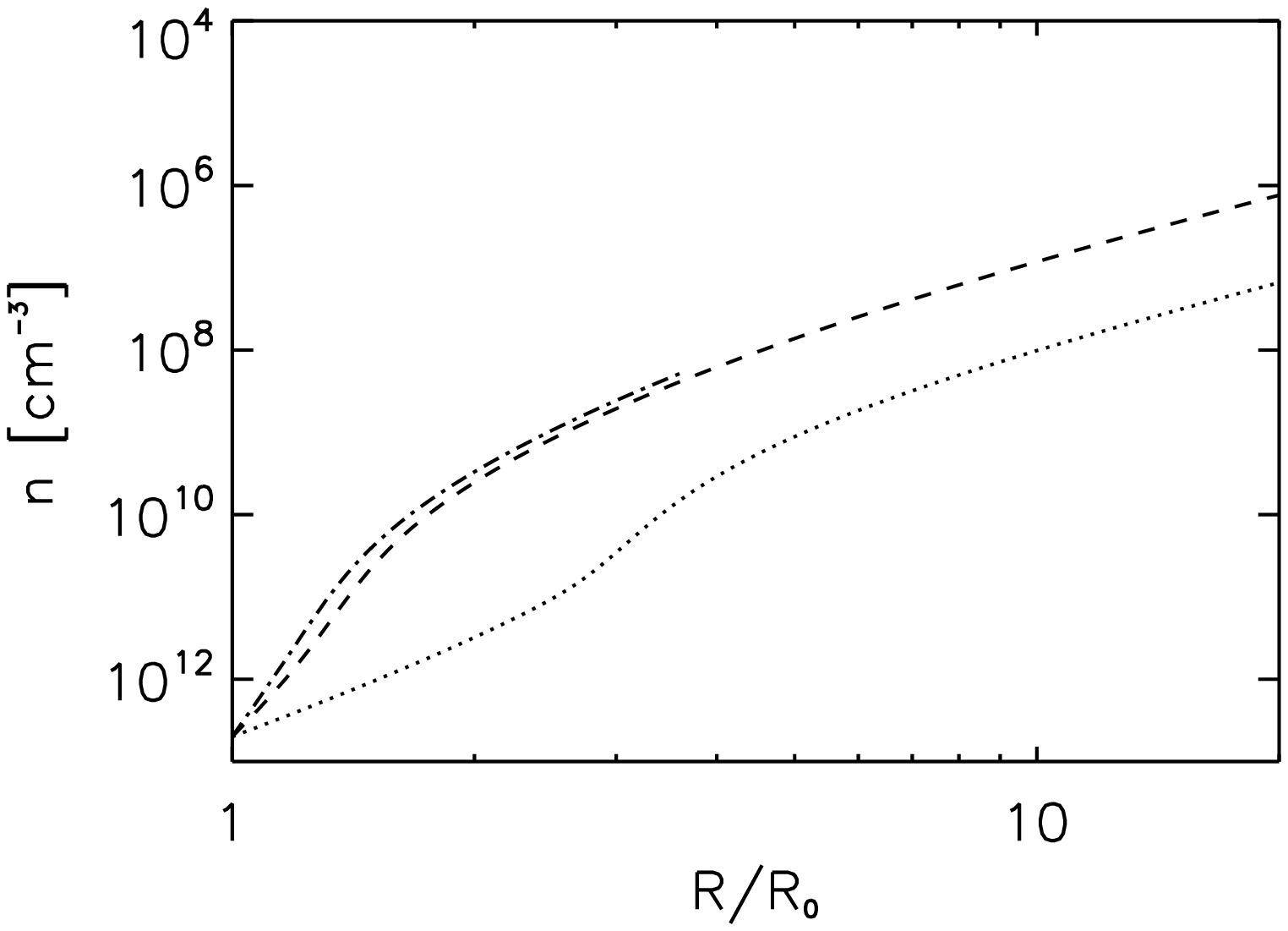}
\includegraphics[width=0.9\columnwidth]{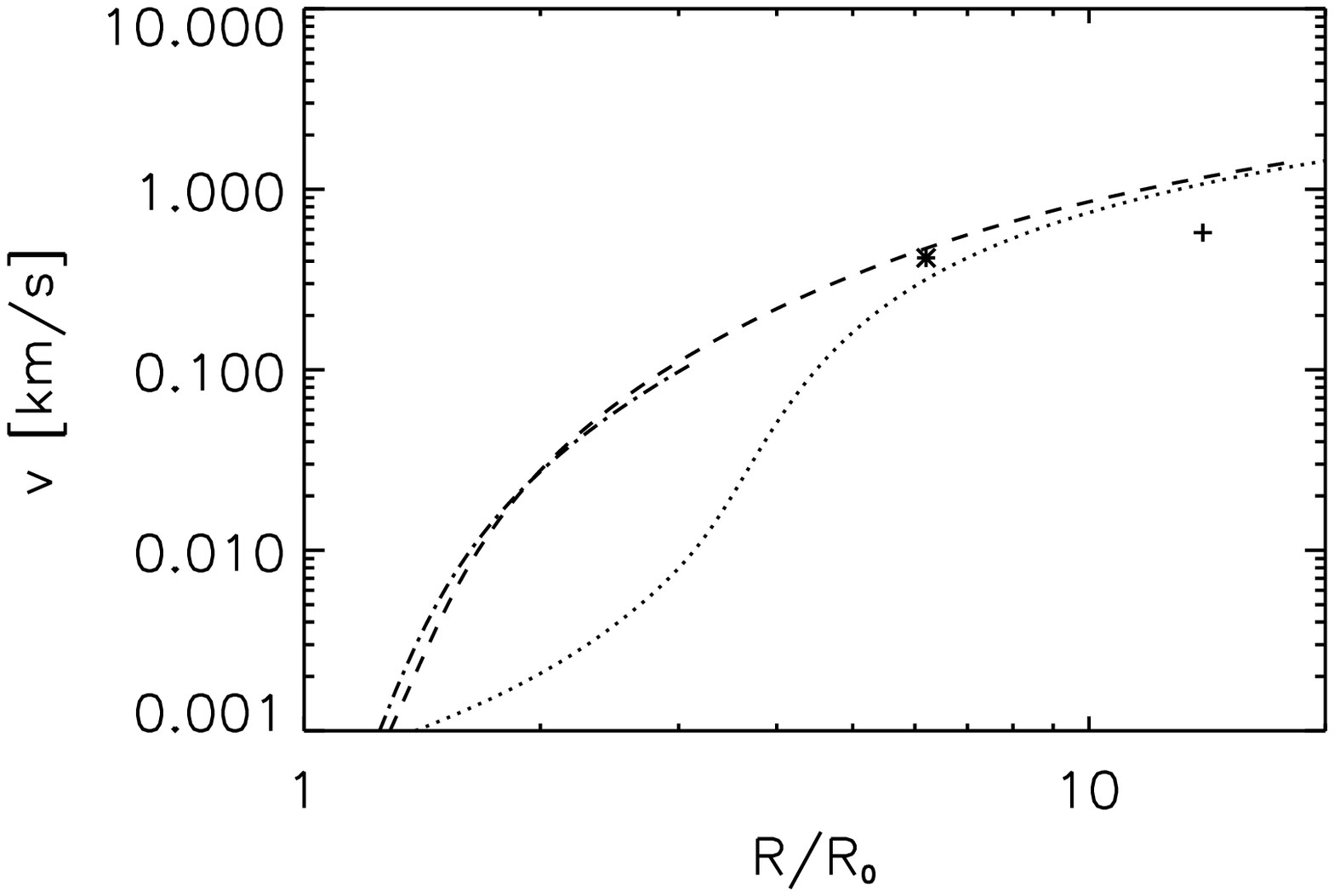}
\includegraphics[width=0.9\columnwidth]{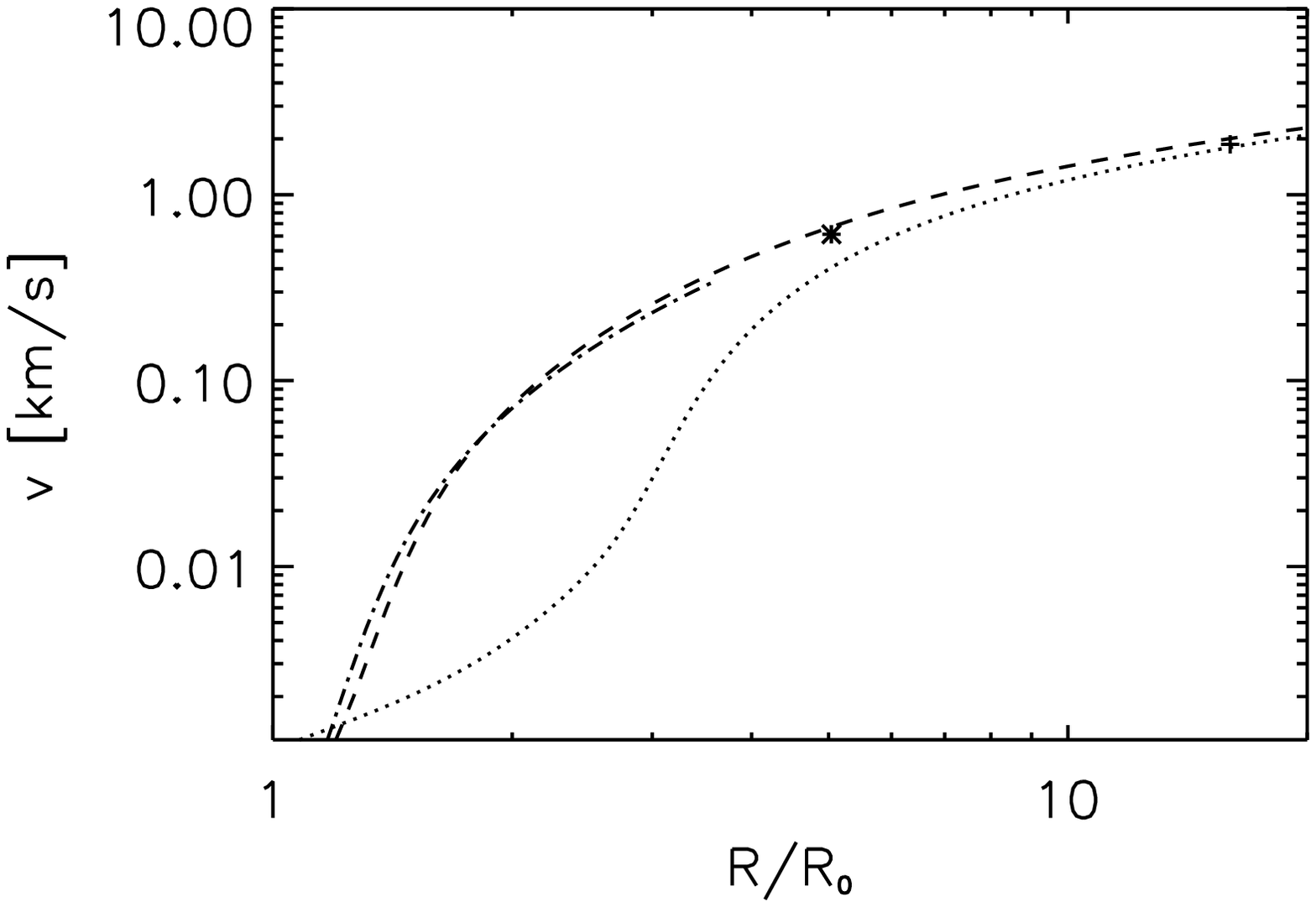}
\caption{From top to bottom profiles of the
XUV volume heating rates, temperatures, densities and velocities of hydrogen envelopes
that are exposed to 100 times higher XUV fluxes compared to today's solar value at 1 AU,
of protoplanets with 0.1$M_{\oplus}$ (dotted lines), 0.5$M_{\oplus}$ (dashed-dotted lines),
and 1$M_{\oplus}$ (dashed lines) core masses
and $\eta$ of 15\% (left column) and 40\% (right column). $z_0$ is assumed to be
0.15$R_{\oplus}$ for the shown protoplanets.
$\ast$ shows the transonic point $R_{\rm s}$ and $+$ $R_{\rm exo}$ for the 1$M_{\oplus}$. In all the
other cases $R_{\rm s} < R_{\rm c}$ (see Table 4).}
\end{center}
\end{figure*}
\begin{figure*}
\begin{center}
\includegraphics[width=0.9\columnwidth]{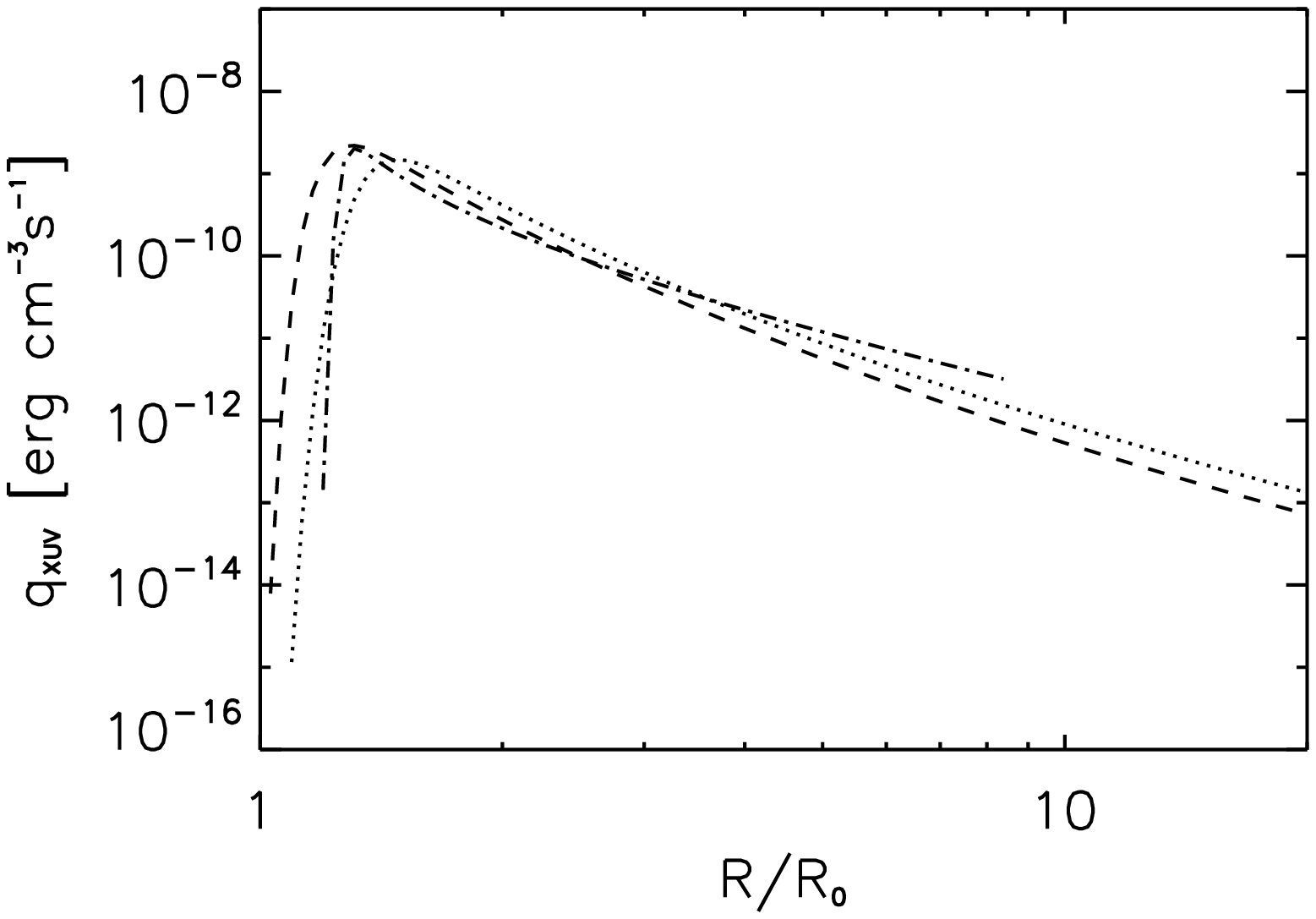}
\includegraphics[width=0.9\columnwidth]{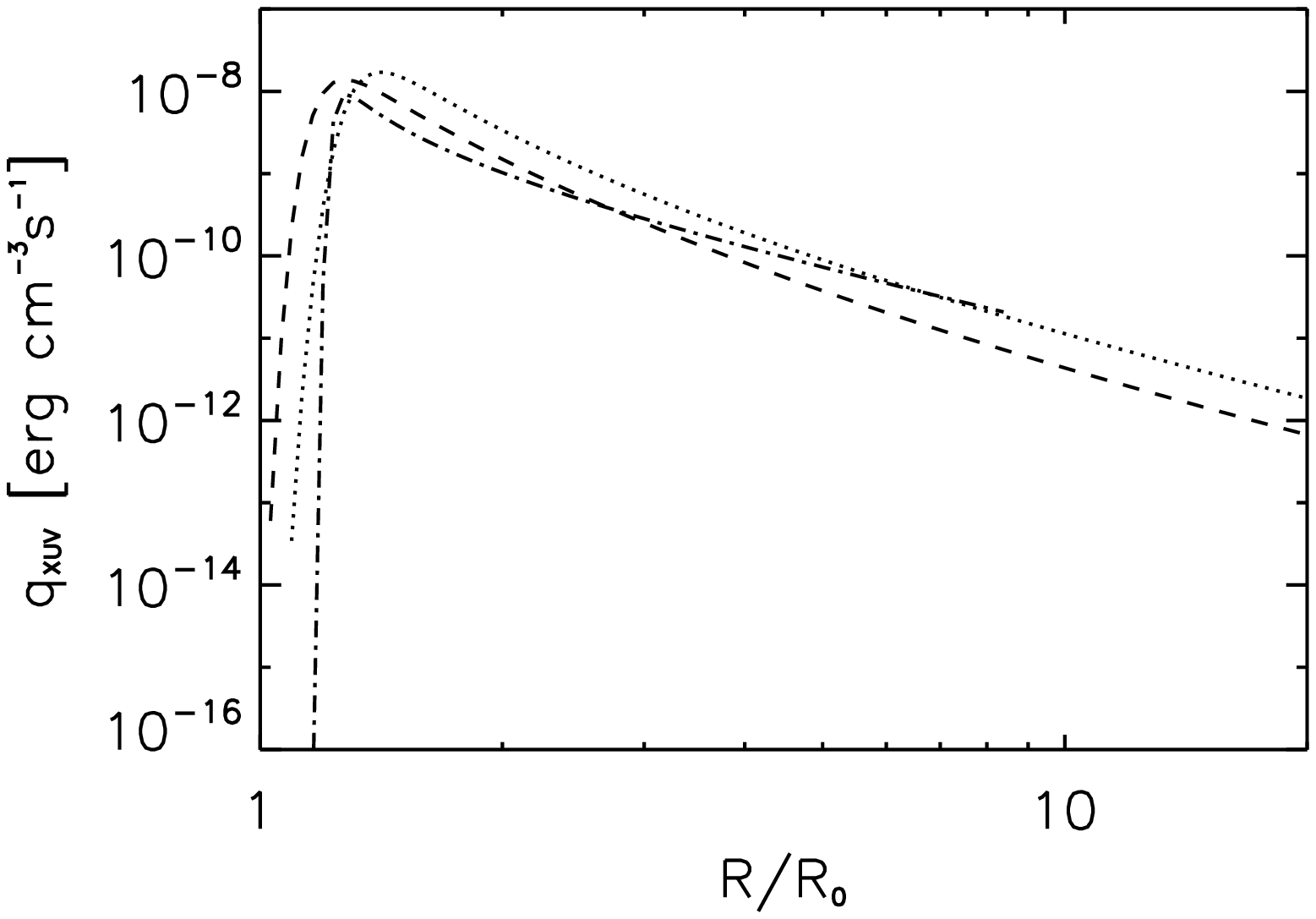}
\includegraphics[width=0.9\columnwidth]{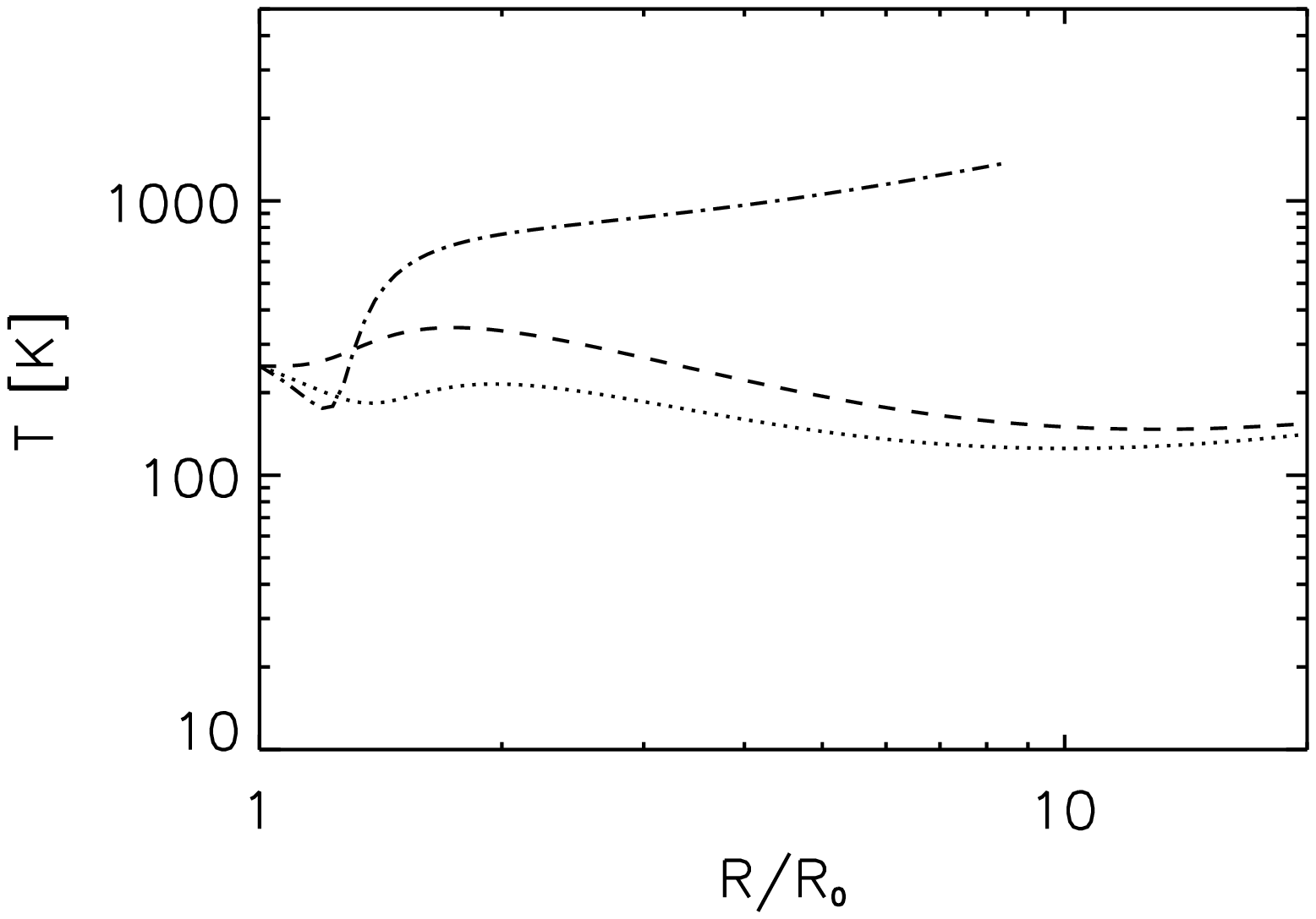}
\includegraphics[width=0.9\columnwidth]{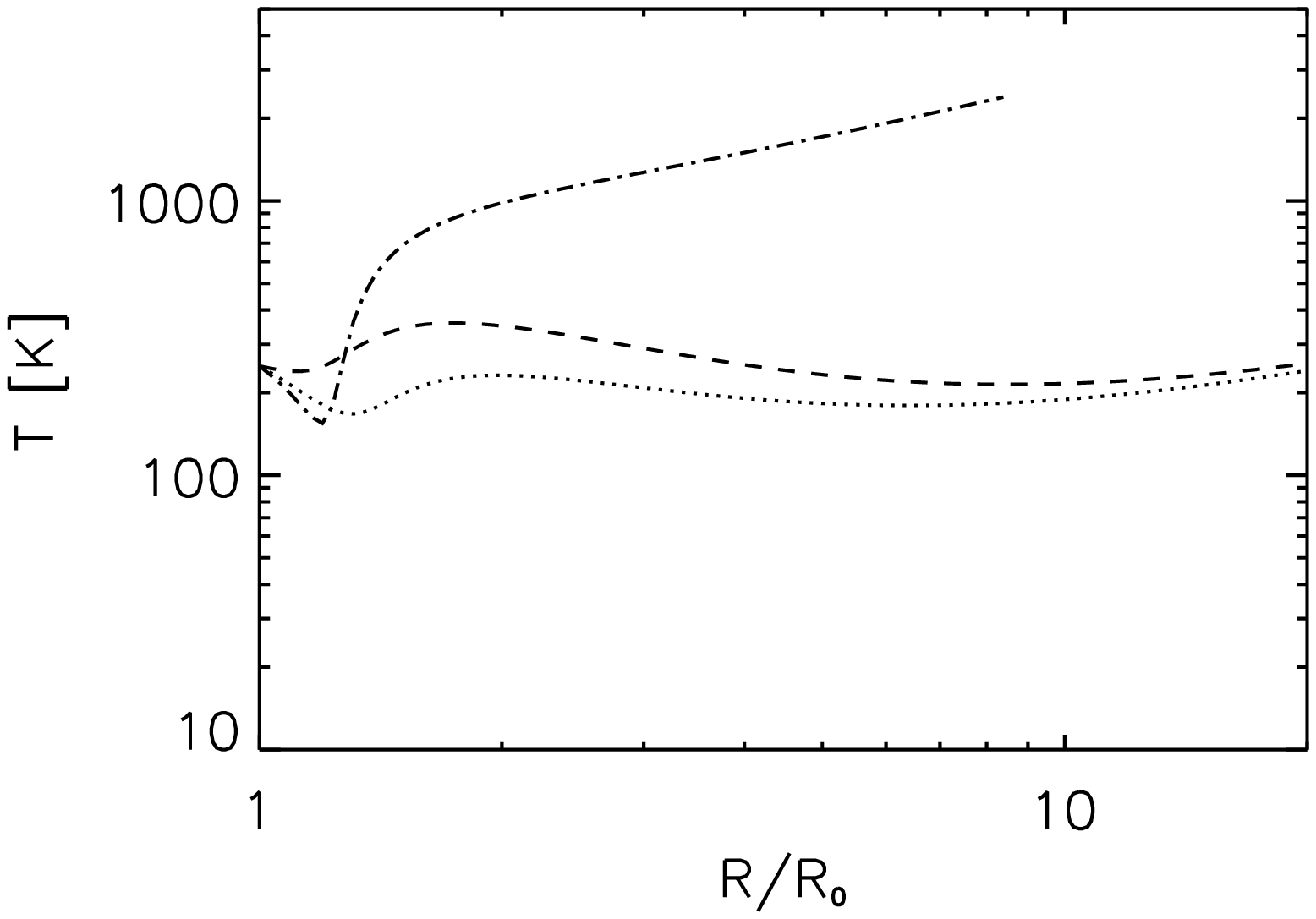}
\includegraphics[width=0.9\columnwidth]{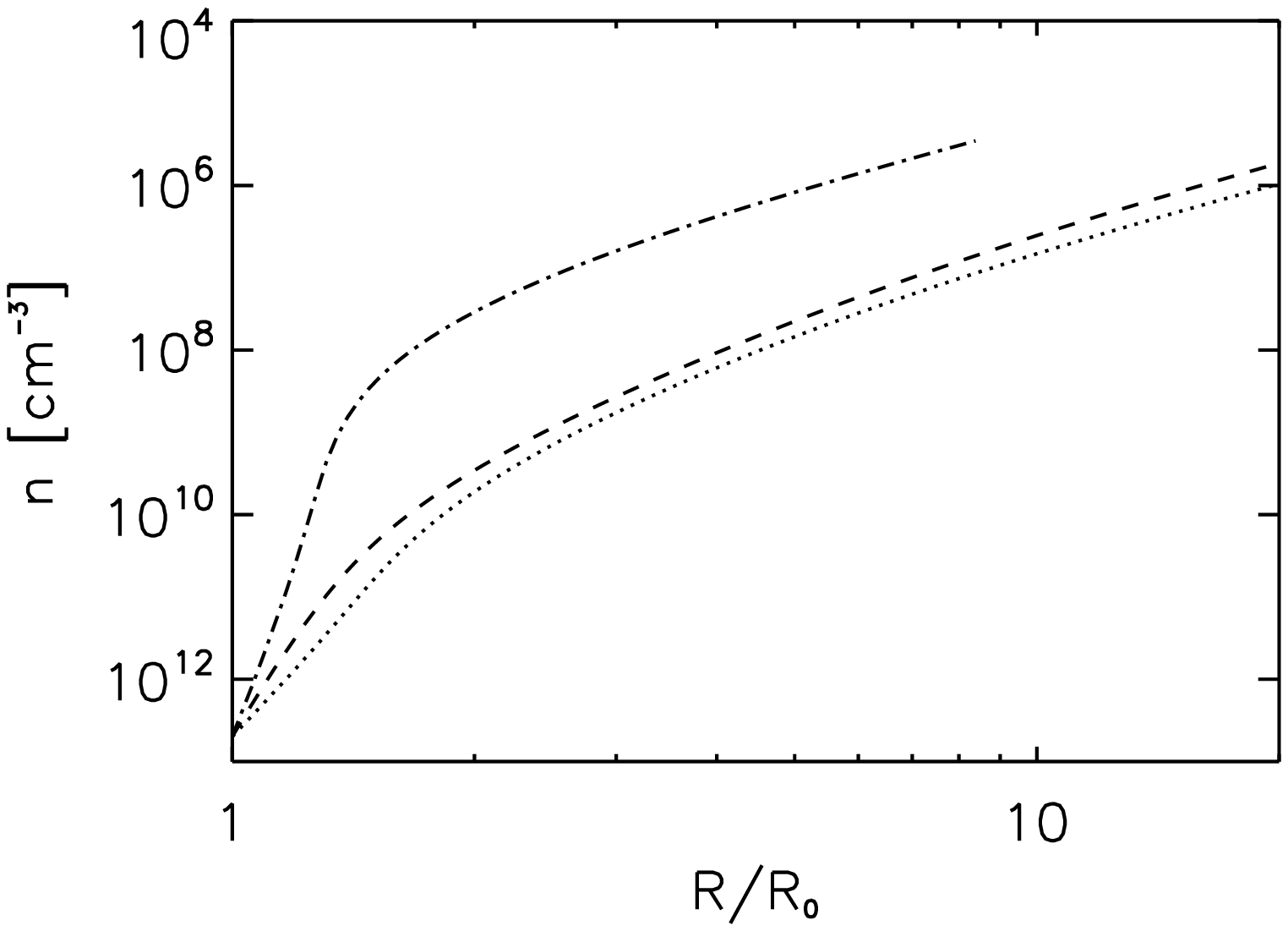}
\includegraphics[width=0.9\columnwidth]{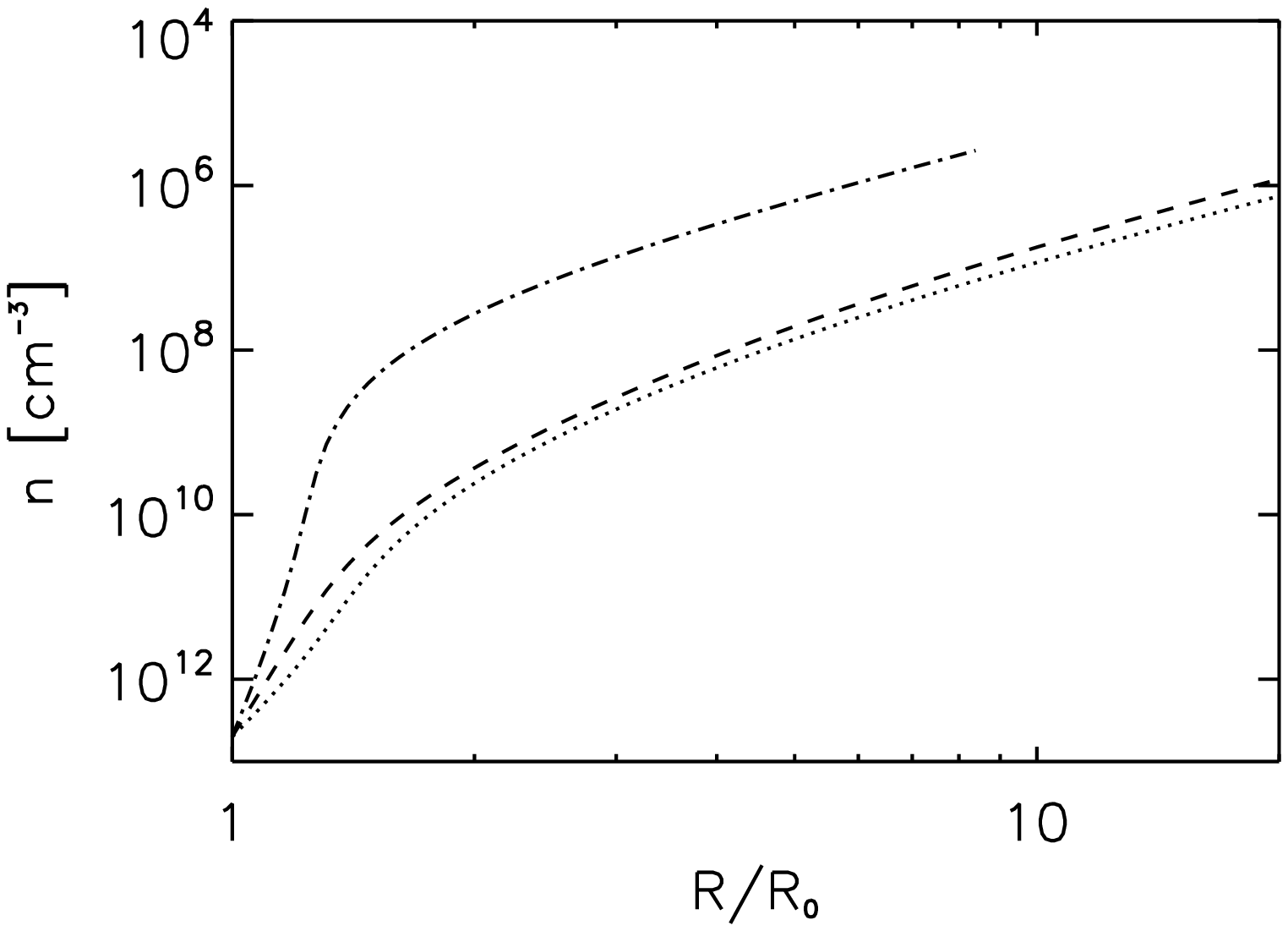}
\includegraphics[width=0.9\columnwidth]{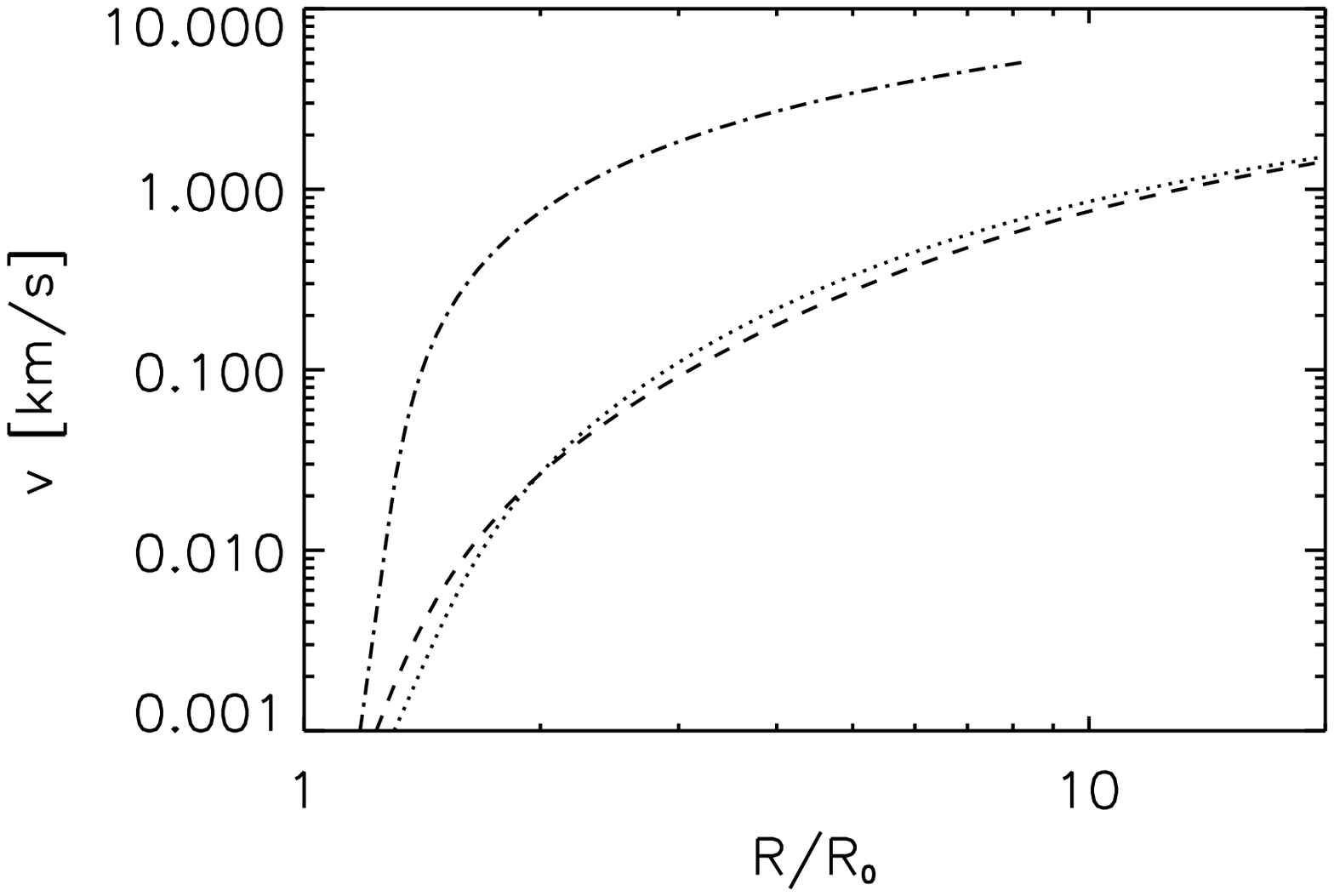}
\includegraphics[width=0.9\columnwidth]{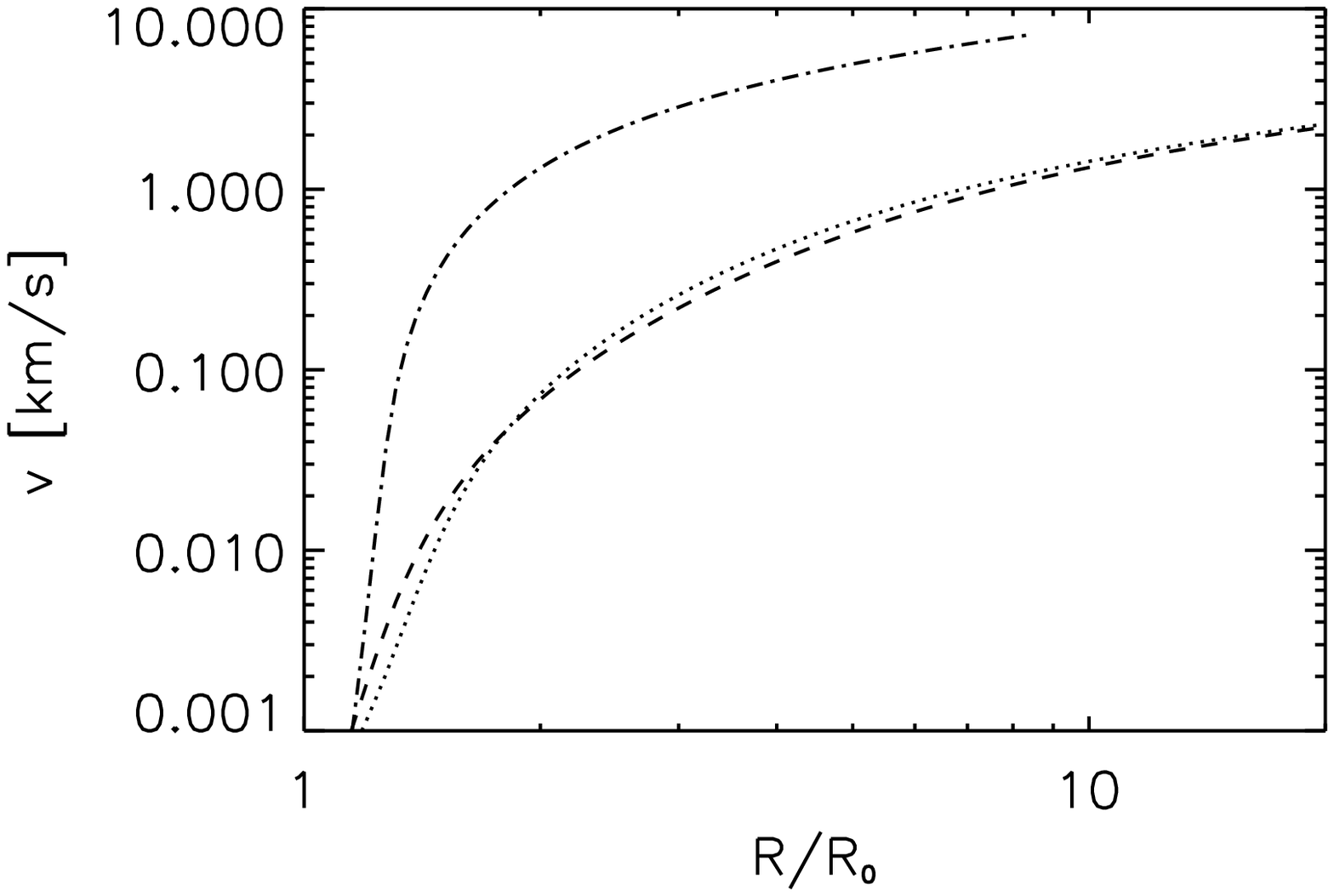}
\caption{From top to bottom profiles of the
XUV volume heating rates, temperatures, densities and velocities of hydrogen envelopes
that are exposed to 100 times higher XUV fluxes compared to today's solar value at 1 AU,
of protoplanets with 2$M_{\oplus}$ (dotted lines), 3$M_{\oplus}$ (dashed-dotted lines),
and 5$M_{\oplus}$ (dashed lines) core masses and $\eta$ of 15\% (left column) and 40\% (right column).
$z_0$ is assumed to be 4$R_{\oplus}$ for all three protoplanets, corresponding to dense hydrogen envelopes $f_{\rm env}$ of 0.1.
In all cases $R_{\rm s} < R_{\rm c}$ (see Table 4).}
\end{center}
\end{figure*}
The incoming XUV flux, which heats the upper atmosphere, decreases
due to absorption near the mesopause through dissociation and
ionization of H$_2$ molecules. We assume that atomic hydrogen is the
dominant species in the upper atmosphere. The XUV volume heating rate $q_{\rm
XUV}$ can then be written in spherical coordinates as
\begin{eqnarray}
q_{\rm XUV}(R,\theta)=\eta\sigma_{\rm XUV}n(R)J_{\rm XUV_0}\nonumber\\
exp\left[-\int_{R cos(\theta)}^{\infty}\sigma_{\rm XUV} n \left(\sqrt{s^2 + R^2 sin^2\theta}\right)ds\right],
\end{eqnarray}
with $J_{\rm XUV_0}$ the XUV flux outside the atmosphere at 1 AU
but enhanced by a factor 100 according to Fig. 1. $\sigma_{\rm
XUV}$ is the photoabsorption cross section of $\sim 5\times
10^{-18}$ cm$^{2}$, which is in agreement with experimental data
obtained by Beyont \& Cairns (1965).

The lower boundary of our simulation domain is chosen to be
located at the base of the thermosphere, $R_0$, as illustrated in
Fig. 2b. This choice is justified by the fact that at its base the
thermosphere is optically thick for the XUV radiation and thus the
bulk of the XUV photons is absorbed above this level and very
little penetrates below it. Thus, $R_0$ can be seen as a natural
boundary between the lower atmosphere
(troposphere-stratosphere-mesosphere) and the upper atmosphere
(thermosphere-exosphere). The absorption of the XUV radiation in
the thermosphere results in ionization, dissociation, heating,
expansion and, hence, in thermal escape of the upper atmosphere.
Because of the high XUV flux and its related ionization and dissociation processes, the escaping
form of hydrogen from a protoatmosphere should mainly be H atoms,
but not H$_2$ molecules. The value of the number density $n_0$ at
the base of the thermosphere is $\sim 5\times 10^{12}$--$10^{13}$
cm$^{-3}$ (e.g., Kasting \& Pollack 1983; Tian et al. 2005; Erkaev
et al. 2013a; Lammer et al. 2013a) and can never be arbitrarily
increased or decreased by as much as an order of magnitude, even
if the surface pressure on a planet varies during its life time by
many orders of magnitude. The reason for this is that the value of
$n_0$ is strictly determined by the XUV absorption optical depth
of the thermosphere.

Below $z_0 = R_0 - R_{\rm pl}$ only a minor fraction of the short
wavelengths radiation can be absorbed, which results in negligible
gas heating compared to the heating of the thermosphere above. The
positive vertical temperature gradient in the lower thermosphere
results in a downward conduction flux of the thermal energy
deposited by solar XUV photons to the mesopause region where this
energy can be efficiently radiated back to space if IR-radiating
molecules like CO$_2$, NO, O$_3$, H$_3^+$, etc., are available.
The IR-cooling rate is the largest near the mesopause, and rapidly
decreases at higher altitudes. As the XUV heating behaves
differently from the IR-cooling and rapidly decreases with a
decreasing altitude towards the mesopause, the combined effect of
the reduced XUV heating and enhanced IR-cooling in the lower
thermosphere results in the temperature minimum $T_0$ at the
mesopause. For planetary atmospheres that are in long-term
radiative equilibrium, $T_0$ is quite close to the so-called
planetary skin temperature, which corresponds to $T_{\rm eff}\sim
T_{\rm eq}\approx 200-250$ K inside the HZ of a
Sun-like star. We point out that an uncertainty of $\pm$25 K has a
negligible effect in the modelled escape rates.

At closer orbital distances, which result in a hotter environment
of the planet's host star and higher surface temperature, the
mesopause level simply rises to a higher altitude where the base
pressure, $P_0$, retains the same constant value as in a less hot
and dense atmosphere at 1 AU. Because it is not known if and which
amounts of IR-cooling molecules are available in the captured
hydrogen envelopes, we study the atmospheric response and thermal
escape by introducing two limiting values of the heating
efficiency $\eta$ which is the fraction of the absorbed XUV
radiation that is transformed into thermal energy. Namely, we use
a lower value for $\eta$ of 15\% (Chassefi\`{e}re 1996a; 1996b;
Yelle 2004; Lammer et al. 2009a; Leitzinger et al. 2011; Erkaev et
al. 2013a) and a higher $\eta$ value of 40\% (Yelle 2004; Penz et
al. 2008; Erkaev et al. 2013a; Koskinen et al. 2013; Lammer et al.
2013a).

Because the rocky core of a protoplanet is surrounded by a dense
hydrogen envelope, with hot surface temperatures that
are related to a greenhouse effect and/or impacts although
after the main accretion phase,
the mesopause can be located from several
hundreds to thousands of kilometers or even up to several Earth
radii $R_{\oplus}$ above the protoplanet's core radius (Wuchterl 2010;
Mordasini et al. 2012; Batygin and Stevenson 2013).
Mordasini et al. (2012) modelled the
total radii of low-mass planets of nebula-based hydrogen envelopes
as a function of a core radius for various envelope mass fractions
$f_{\rm env}$. The base of the thermosphere altitude $z_0$, and
the corresponding radius, $R_0$, where the atmosphere temperature
reaches the equilibrium temperature at the test planet's orbital
location, are estimated from the results of Mordasini et al.
(2012). However, according to their study (rocky cores: Fig. 3 right panel of
Mordasini et al. 2012) the $R_0$ values as a function of $f_{\rm
env}$ for low mass bodies should be considered only as rough
estimates (Mordasini \& Alibert 2013; private communication).
Therefore, we assume in agreement with the modelled radii by Mordasini et al. (2012)
for thin hydrogen envelopes with $f_{\rm env}
\leq 0.001$ a value for $z_0$ of 0.15$R_{\oplus}$, for envelopes
$f_{\rm env}\sim 0.01$ $z_0$ of $1R_{\oplus}$, and for
$f_{\rm env}\sim 0.1$ a larger height for $z_0$ of $4R_{\oplus}$.

The upper boundary in our simulation domain is chosen at 70$R_{\rm
pl}$, but according to molecular-kinetic simulations
by Johnson et al. (2013) modelled atmospheric profiles
of transonic outflow should be considered as accurate only
up to a Knudsen number $Kn$ of $\approx 0.1$, that is the ratio of the mean free path
to the corresponding scale height. Above this critical distance $R_{\rm c}$ up to the
exobase level $R_{\rm exo}$ ($Kn=1$) there is a transition region between the hydrodynamic and kinetic domain,
in where the hydrodynamic profiles begin to deviate from gas kinetic models.
In some cases, when $R_{\rm exo}$ is located above 70$R_{pl}$ analytical asymptotics to
extend the numerical solution have been used. In the supersonic region of
such cases we have reduced the full system of equations to two first order
ordinary differential equations for the velocity and entropy.
Then we integrate these ordinary equations along the radius.
By comparing the results for the density one can see that the
obtained power law is a good approximation for the density behaviour.
It should also be noted that Erkaev et al. (2013a) found that in spite of differences in the resulting
atmospheric profiles, in this transition region the escape rates are more or less similar.

Because of the findings of Johnson et al. (2013) we consider that all outward
flowing hydrogen atoms escape by blow-off from the protoplanet only if
the kinetic energy of the hydrodynamically outward flowing atmosphere
is larger than the potential energy in the gravity field of the planet at $R_c$.
If the dynamically expanding hydrogen gas does not reach
the escape velocity $v_{\rm esc}$ at $R_{\rm c}$,
then blow-off conditions are not established and the escaping hydrogen atoms
are calculated by using an escape flux based on a shifted Maxwellian velocity distribution that is modified by the radial velocity, obtained from the hydrodynamic model (e.g., Tian et al., 2008a; 2008b; Volkov et al., 2011; Erkaev et al., 2013a).
In such cases the escape rates are higher compared to the classical Jeans escape, where the atoms
have velocities according to a Maxwell distribution, but lower than
hydrodynamic blow-off.
\subsection{Hydrogen loss}
Table 4 shows the input parameters, including the hydrogen
envelope related lower boundary altitude $z_0$, the distance where
the outward expanding bulk atmosphere reaches the transonic point
$R_{\rm s}$, the corresponding Knudsen numbers $Kn_{\rm s}$,
the critical distance $R_{\rm c}$, where
$Kn\sim$0.1, the exobase level $R_{\rm exo}$ where $Kn=1$,
and the thermal escape rate $L_{\rm th}$, as well as
the total atmosphere loss $L_{\rm \Delta t}$ integrated during the XUV
activity saturation period $\Delta t = 90$ Myr, after the nebula
dissipated. As discussed in Sect. 2 and shown in Fig. 1,
after this phase, the XUV flux decreases during the following
$\sim$400 Myr from an enhancement factor of $\sim$100 to $\sim$5.
As it was shown by Erkaev et al. (2013a), the decreasing XUV flux
yields lower escape rates and hence less loss. Depending on the
assumed heating efficiency $\eta$ and the planet's gravity,
especially `super-Earths' at 1 AU in G-star habitable zones will
not lose much more hydrogen than $\sim$1.5--7 EO$_{\rm H}$,
during their remaining lifetime (Erkaev et al. 2013a; Kislyakova
et al. 2013). For this reason we focus in this parameter study
only on the first 100 Myr, when the XUV flux and related mass loss can be
considered as most efficient.
\begin{figure*}
\begin{center}
\includegraphics[width=0.9\columnwidth]{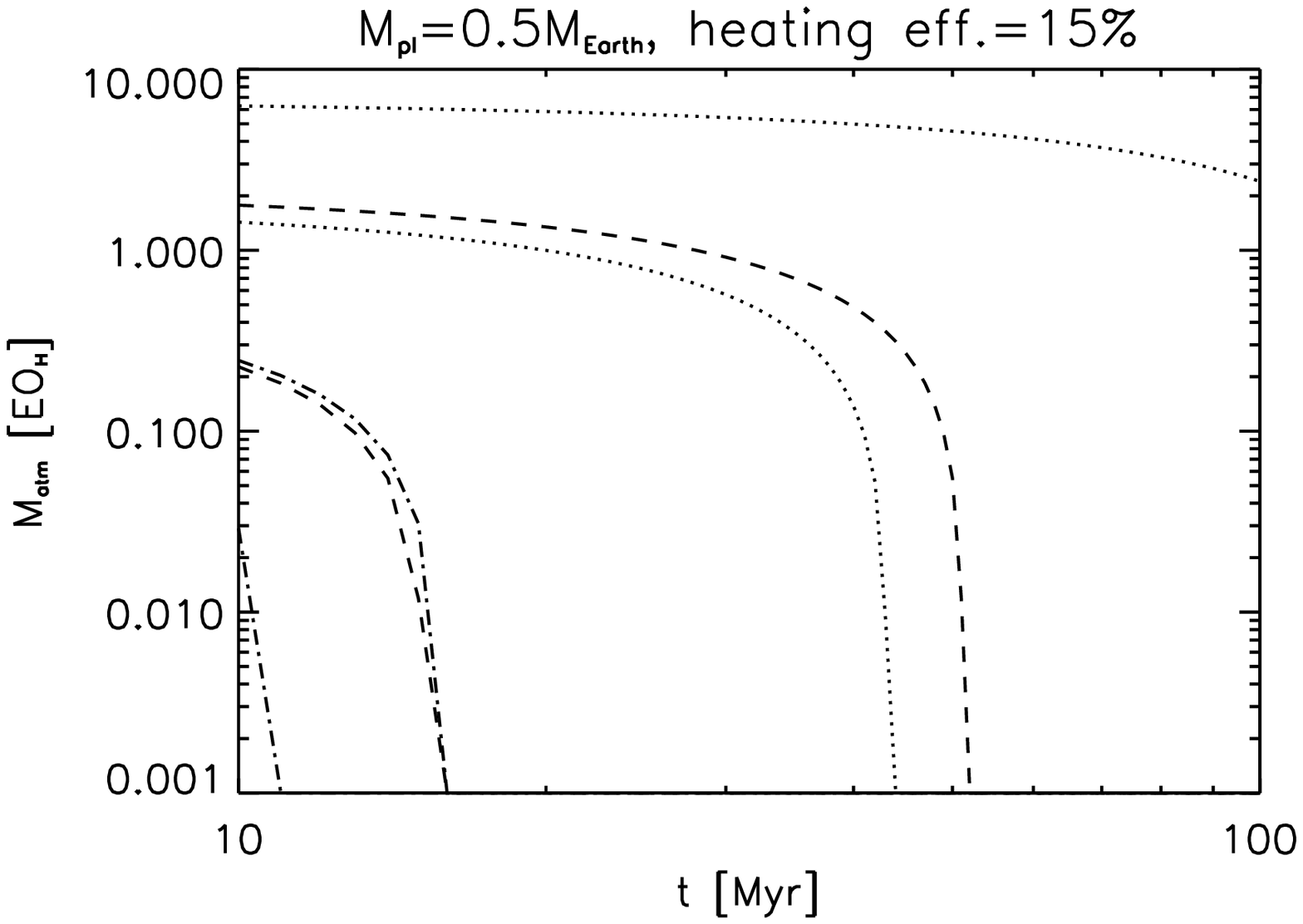}
\includegraphics[width=0.9\columnwidth]{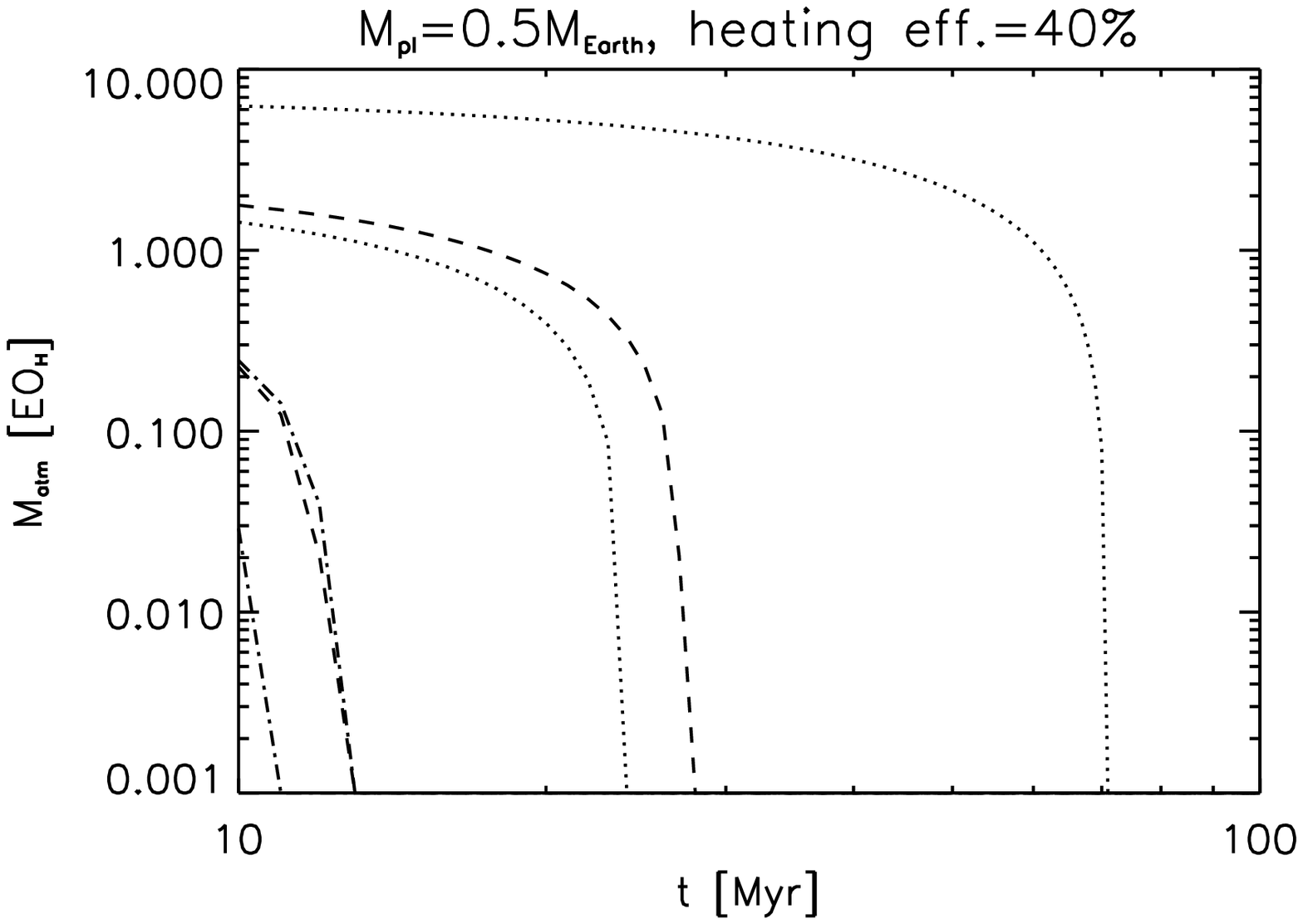}
\includegraphics[width=0.9\columnwidth]{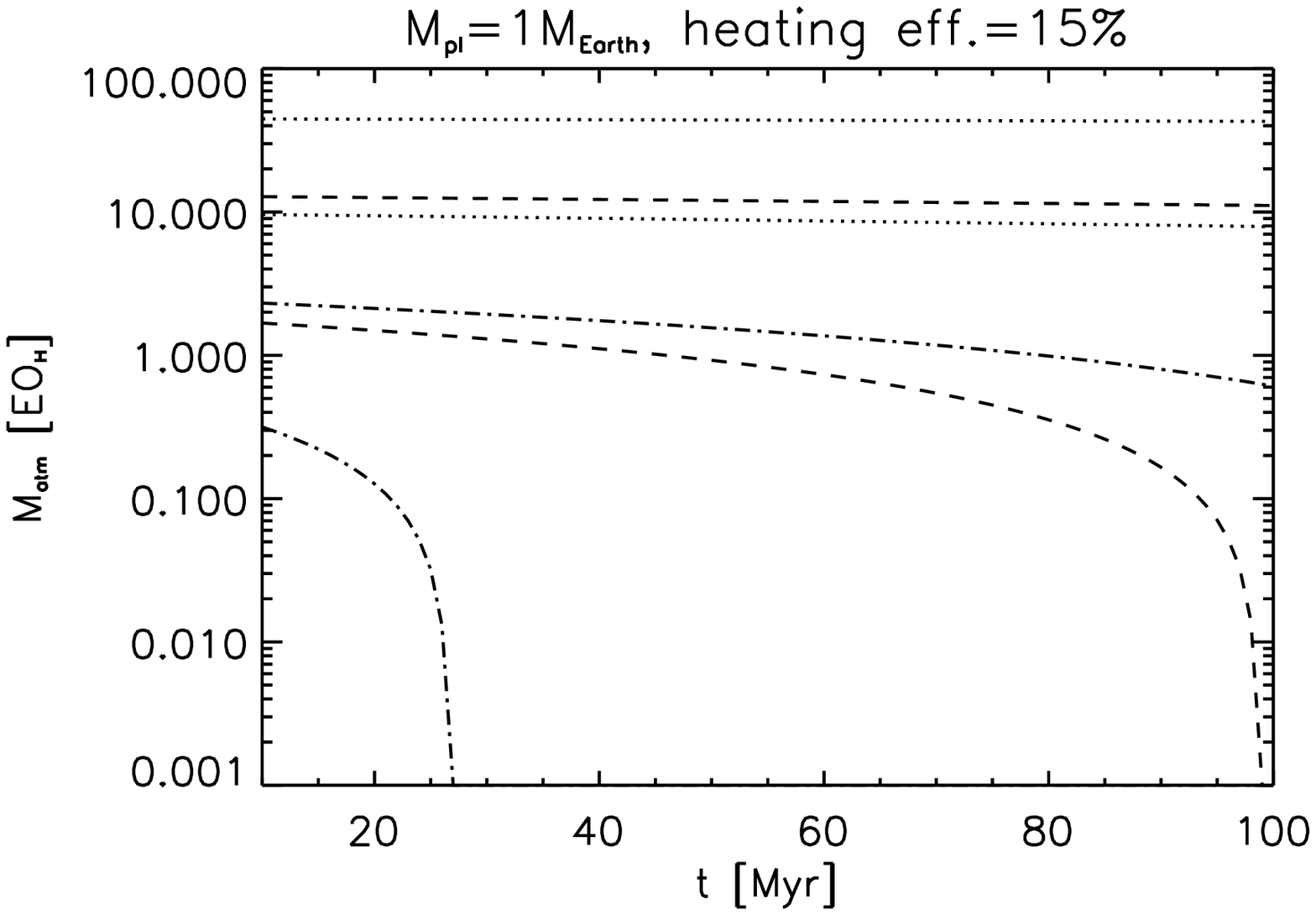}
\includegraphics[width=0.9\columnwidth]{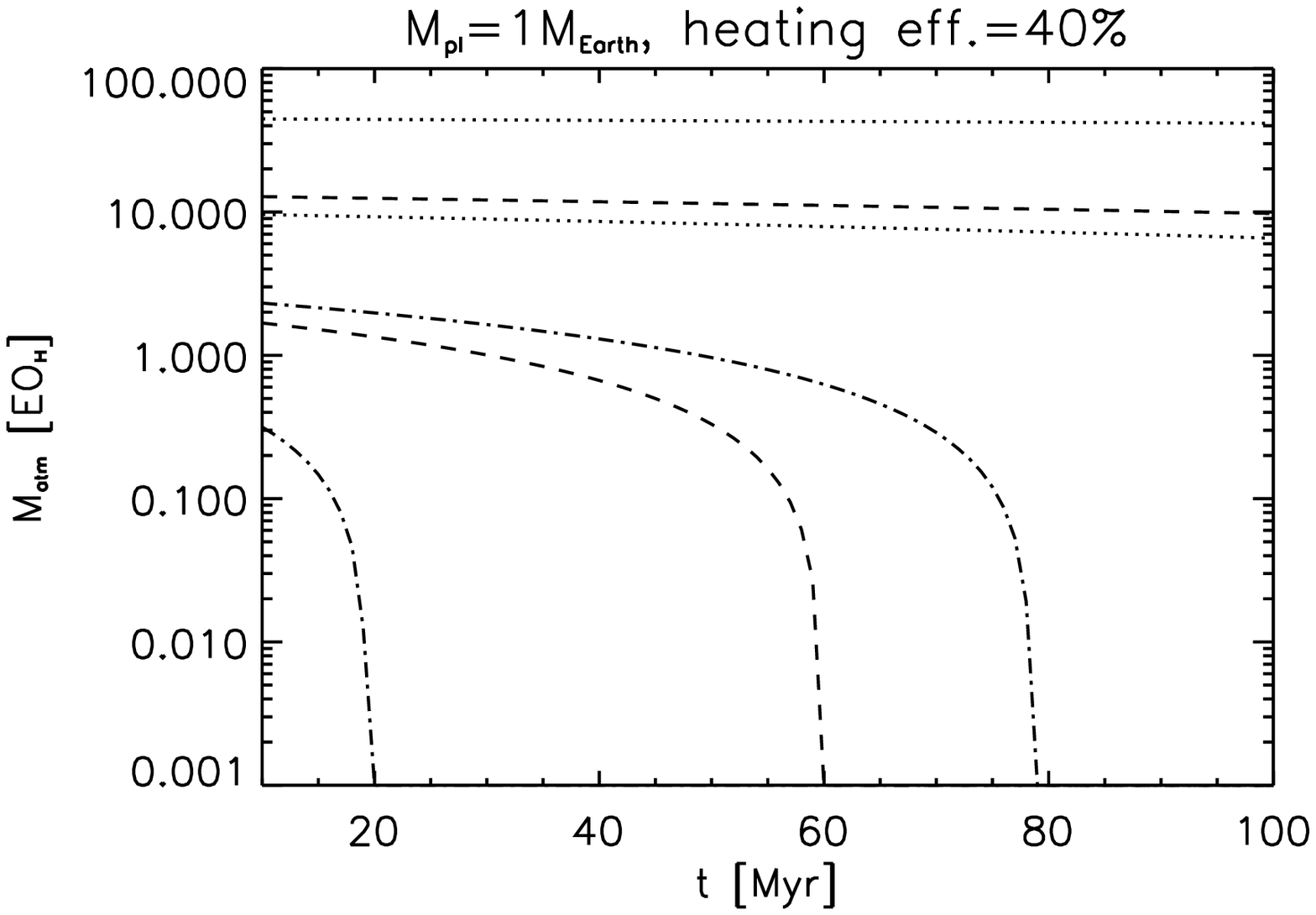}
\includegraphics[width=0.9\columnwidth]{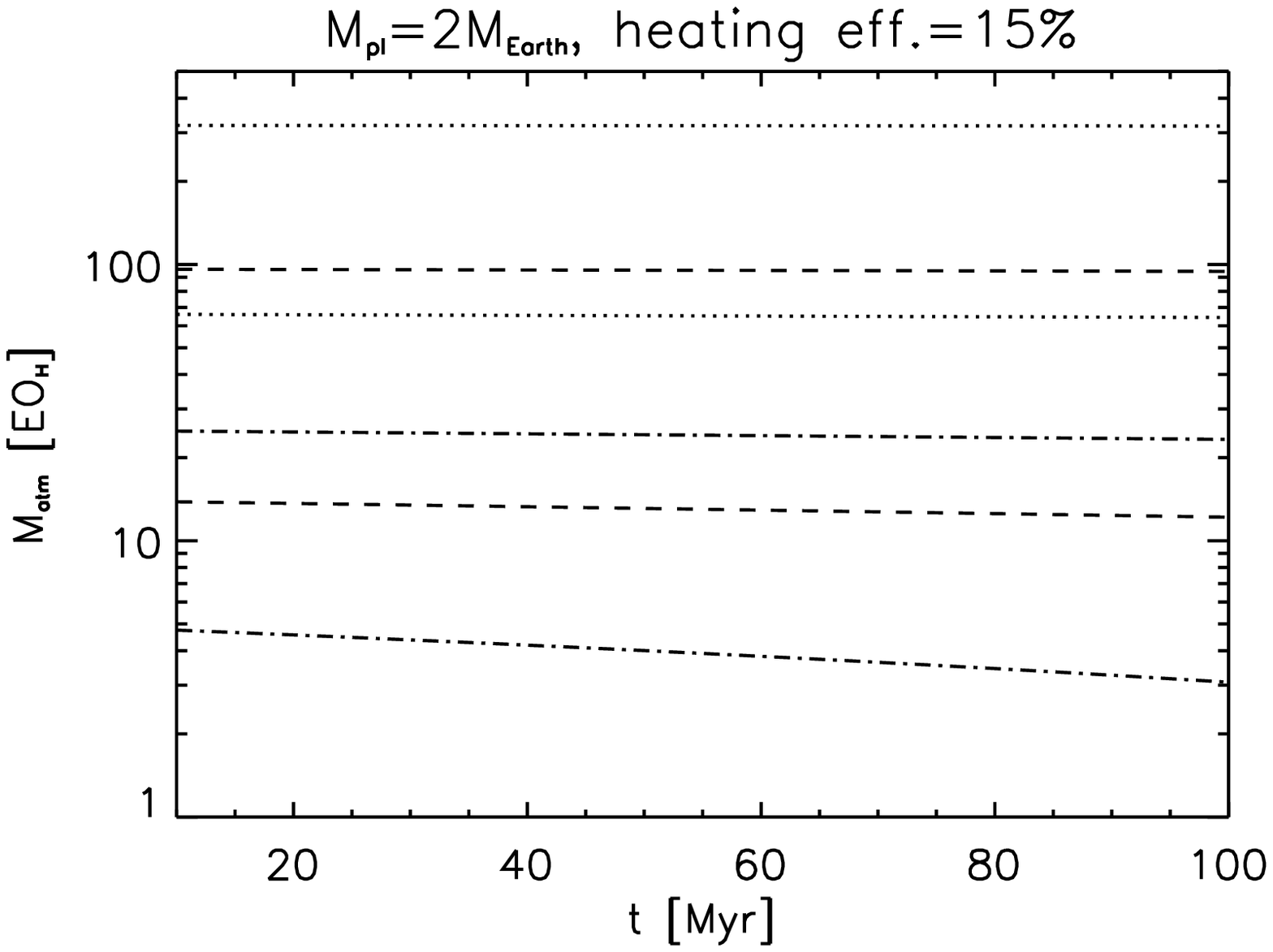}
\includegraphics[width=0.9\columnwidth]{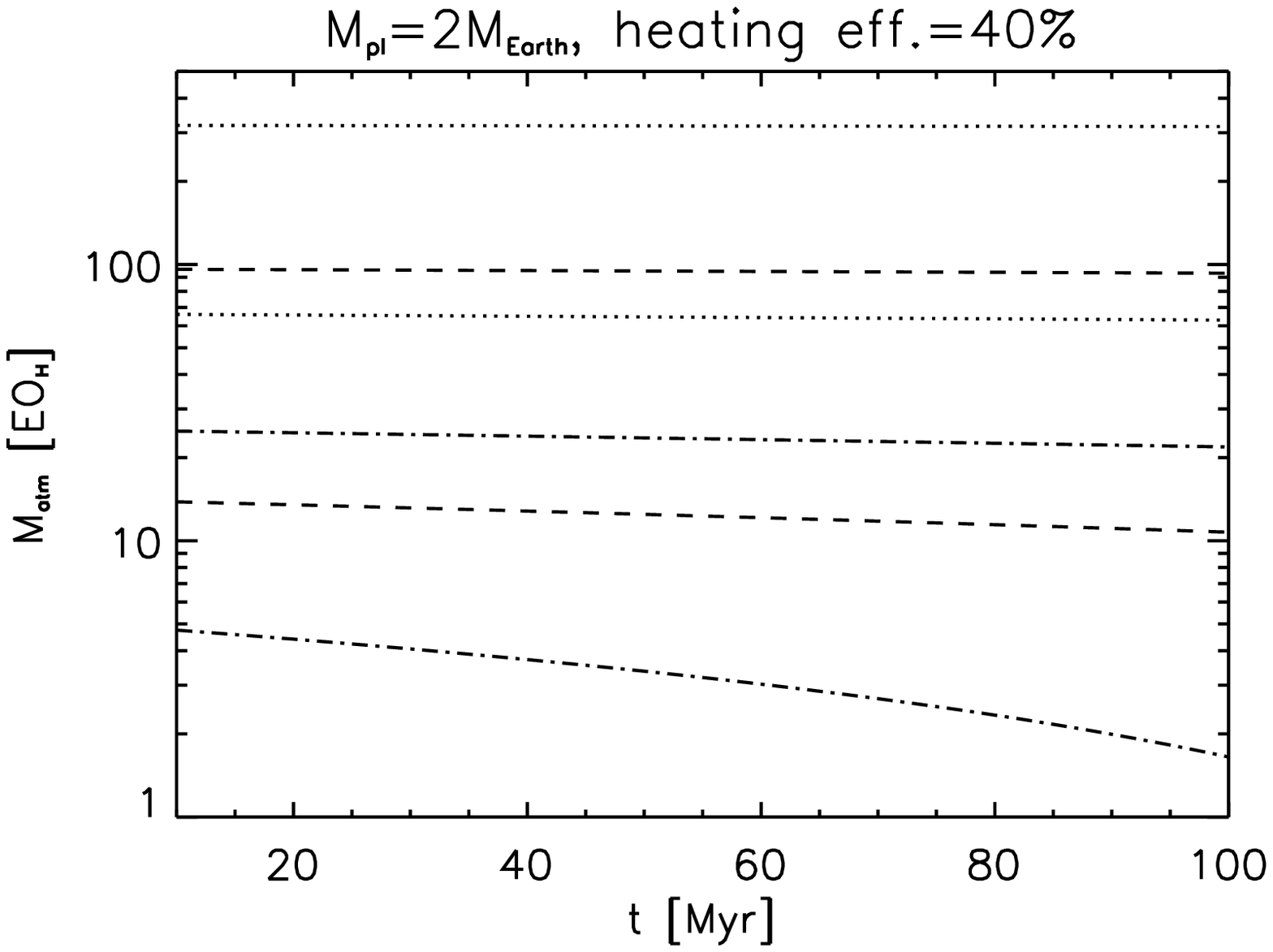}
\caption{Evolution of captured hydrogen envelopes during the
activity saturated phase of a solar like G star for a protoplanet
with 0.5$M_{\rm \oplus}$ (upper panels), 1$M_{\rm \oplus}$ (middle
panels) and for a ``super-Earth'' with 2$M_{\rm \oplus}$ (lower
panels) for heating efficiencies $\eta$ of 15\% (left column) and
40\% (right column). The dotted lines correspond to a nebula dust
depletion factor $f$ of 0.001, the dashed lines to $f$=0.01 and
the dashed-dotted lines to $f$=0.1. The upper lines which yield a
less effective loss are related to a relative accretion rate
$\dot{M}_{\rm acc}/M_{\rm pl}$ of $10^{-5}$ yr$^{-1}$, while the lower lines
correspond to a $\dot{M}_{\rm acc}/M_{\rm pl}$ of $10^{-6}$ yr$^{-1}$.
Lower relative accretion rates and corresponding luminosities as shown
in Tables 1 to 3 yield more massive hydrogen envelopes, that can
not be lost during a protoplanet's lifetime.}
\end{center}
\end{figure*}
\begin{table*}
\renewcommand{\baselinestretch}{1}
\caption{Input parameters, $f_{\rm env}$ and related $z_0$ the height
of the radius $R_0$ above the core surface $R_{\rm pl}$, the
location of the transonic point $R_{\rm s}$, the corresponding
Knudsen number $Kn_{\rm s}$ at this location, the critical radius
$R_{\rm c}$ where $Kn$ reaches 0.1, and the exobase distance $R_{\rm exo}$,
as well as the thermal escape rate of hydrogen atoms
over 3$\pi$ steradian in units of $s^{-1}$
and total loss during $\Delta t$ of 90 Myr in units of EO$_{\rm H}$ for
protoplanet core masses of 0.1$M_{\rm \oplus}$, 0.5$M_{\rm
\oplus}$, 1.0$M_{\rm \oplus}$, 2.0$M_{\rm \oplus}$, 3.0$M_{\rm
\oplus}$, and 5$M_{\rm \oplus}$.}
\begin{center}
\begin{tabular}{cccccccccc}
Core mass [$M_{\rm \oplus}$] & $f_{\rm env}$ & $\eta$ [\%] & $z_0$ [$R_{\rm \oplus}$] & $R_{\rm s}$ [$R_{0}$] & $Kn_{\rm s}$ & $R_{\rm c}$ [$R_0$]& $R_{\rm exo}$ [$R_0$]& $L_{\rm th}$ [s$^{-1}$] & $L_{\rm \Delta t}$ [EO$_{\rm H}$] \\\hline\hline
0.1  &  0.001 &15 & 0.15 &  16.7   & 0.006 & 180   &500  & $ 1.3\times10^{33}$   & 40\\
0.1  &  0.001 &40 & 0.15 &  12.8  & 0.004  & 200  &880  & $1.8\times10^{33}$   & 55.7\\ \hline
0.5  &  0.001&15 & 0.15 & 20.0   & 0.125 &  18  &80  & $1.25\times10^{32}$& 3.9\\
0.5  &  0.001&40 & 0.15 & 14.0 & 0.05 & 25  &170  & $3.0\times10^{32}$ & 9.3\\
0.5  &  0.01&15 & 0.5 &  18 & 0.05   & 25  &144  & $3.5\times10^{32}$ & 10.8\\
0.5  &  0.01&40 & 0.5 &  12.4 & 0.02  & 41 &280   & $7.7\times10^{32}$ & 23.8\\ \hline
1  &  0.001&15 & 0.15 & 6.2 & 0.3  & 3.2  &14  &  $5.5\times10^{31}$  & 1.73\\
1  &  0.001&40 & 0.15 & 5.0 & 0.16   & 3.7  &16  & $9.8\times10^{31}$  & 3\\
1  &  0.01&15 & 1 &  20.0 & 0.06   & 32 &220   &$5.6\times10^{32}$   & 17.3\\
1  &  0.01& 40& 1 &  14.0 & 0.02  & 47 &300   & $1.3\times10^{33}$   & 40.3\\ \hline
2  & 0.001& 15 & 0.15  & 7.2 & 0.46 &  3 & 12 &  5.5$\times 10^{31}$ & 1.7 \\
2  & 0.001&  40 & 0.15 & 5.5 & 0.22 &   3.5 &15  &  $1.0\times 10^{32}$ & 3 \\
2  & 0.01 & 15 & 1   & 6 & 0.15  &  4.6 &20  & $2.0\times 10^{32}$ & 6 \\
2  & 0.01 & 40 & 1   & 5 & 0.11    &   5.5 & 26  &  3.7$\times 10^{32}$ & 11.5\\
2  & 0.1 &  15 & 4   & 17.7 & 0.012  &  100 &500  & 6.2$\times 10^{33}$ & 192 \\
2  & 0.1 & 40 & 4   & 12.4 & 0.0045  &  140 &700  &  1.3$\times 10^{34}$ & 403 \\\hline
3  & 0.001 & 15 & 1  & 6.6 & 0.2   & 4.2 &19  & 1.8$\times 10^{32}$ &5.6 \\
3  & 0.001 & 40 & 1  & 5.35 & 0.11  &  5 &23  &  3.5$\times 10^{32}$ & 10.9\\
3  & 0.01 & 15 & 1.35  & 6.4 & 0.16   &  4.9 &22  &  2.6$\times 10^{32}$ & 8 \\
3  & 0.01 & 40 & 1.35  & 5.2 & 0.08   &  5.9 &28  & 5.0$\times 10^{32}$ & 15.5\\
3  & 0.1 & 15 & 4    & 20 & 0.03 &  60  &320 & 3.5$\times 10^{33}$ &108.5\\
3  & 0.1 & 40 & 4    & 14.8 & 0.01  &  90 &480  &  8.4$\times 10^{33}$ & 260\\\hline
5  & 0.001 &  15 & 1.2 & 7.7 & 0.3   &  4.2  &17   &  2.0$\times 10^{32}$ & 6.2\\
5  & 0.001 & 40 & 1.2  & 5.9 & 0.13   &  5  &23  &  4.0$\times 10^{32}$ & 12.4\\
5  & 0.01 & 15 & 1.5   & 7.5 & 0.23   &  4.5  &19  & 2.6$\times 10^{32}$ & 8\\
5  & 0.01 & 40 & 1.5   & 5.7 & 0.11    &  5.5  &25  &  5.2$\times 10^{32}$ & 16.1\\
5  & 0.1 & 15 & 4    & 6 & 0.06   &  8.5  &45  &  1.3$\times 10^{33}$ & 40.3\\
5  & 0.1 & 40 & 4    & 5 & 0.03   &  8.5   &57  &  2.4$\times 10^{33}$ & 74.4\\\hline
\end{tabular}
\end{center}
\normalsize
\end{table*}
One can see from Mordasini et al. (2012), that $z_0$ and $R_0$ move to
higher distances if the protoplanets that are surrounded by a more
massive hydrogen envelope. We consider this aspect also in the
results presented in Table 4. One can see from the escape rates in Table 4
that they can be very high if $z_0$ and its related $R_0$ are large.
But on the other hand one can see from Mordasini et al. (2012), if $R_0$ is
large, then $f_{\rm env}$ is also larger and this compensates the
ratio of the lost hydrogen to the initial content over time,
so that the majority of the captured hydrogen can not be lost
during the planet's lifetime although the escape rates are very
high.

One can also see from Table 4 that the hydrodynamically expanding upper
atmosphere reaches the transonic point $R_{\rm s}$ in 13 cases out
of 28 before or near the critical radius $R_{\rm c}$.
However, in most cases where $R_{\rm s}>R_{\rm c}$ the corresponding
Knudsen number $Kn_{\rm s}$ is only slightly higher than 0.1, but in
all cases it is $<$0.5. Thus, in the majority of cases the transonic
point is reached well below the exobase level $R_{\rm exo}$ where $Kn$ is 1,
and nearly all outward flowing hydrogen atoms will escape from the planet.

The first row in Fig. 3 show the XUV volume heating rate $q_{\rm XUV}$ profiles, corresponding
to $\eta$ of 15\% (left column) and 40\% (right column), as a
function of $R/R_0$
for the 0.1$M_{\oplus}$, 0.5$M_{\oplus}$, and 1$M_{\oplus}$ protoplanets. $f_{\rm env}$
values are related to cases where $z_0=R_0-R_{\rm pl}$ is assumed to be 0.15$R_{\oplus}$
for the `sub-Earths' and Earth-like protoplanets.
The second, third, and fourth rows in Fig. 3 show the corresponding temperature profiles,
density and velocity profiles of the captured hydrogen envelopes for the same `sub-' to Earth-mass
protoplanets.
One can also see that for the Earth-like core $R_{\rm c}$ is reached before the transonic point
$R_{\rm s}$ (marked with $\ast$), but well below the location $R_{\rm exo}$ (marked with $+$)
where $Kn=1$. For all the other cases shown in Fig. 3, $R_{\rm s} < R_{\rm c}$.
One can see from the profiles that core masses around
0.5$M_{\oplus}$ represent a kind of a boundary separating
temperature regimes for low and high mass bodies. Below this mass range,
because of the low gravity, the gas cools and expands to further distances until the
XUV radiation is absorbed. In the case of the low gravity 0.1$M_{\rm \oplus}$ planetary
embryo-type protoplanet no XUV radiation is absorbed near $z_0$ but further out.
For more massive bodies the XUV heating dominates much closer to $z_0$ and adiabatic
cooling reduces the heating only at further distances.

The first, second, third and fourth rows in Fig. 4 show the profiles of the
XUV volume heating rates, temperatures, densities and velocities of the hydrodynamically
outward flowing hydrogen atoms for the `super-Earths' with  2$M_{\oplus}$,
3$M_{\oplus}$, and 5$M_{\oplus}$ core masses for heating efficiencies $\eta$ of 15\% (left column) and
40 \% (right column). The shown `super-Earth' case profiles
correspond to dense hydrogen envelopes $f_{\rm env}$, where $z_0=R_0-R_{\rm pl}$ is
assumed to be 4$R_{\oplus}$. For all cases shown in Fig. 4,
the transonic point $R_{\rm s} < R_{\rm c}$.

If one compares the captured hydrogen envelope masses shown in
Tables 1 to 3 with the losses during the most efficient G star XUV
flux period $\Delta t$ after the nebula dissipated, then one finds
that protoplanetary cores with masses between 0.1$M_{\rm \oplus}$
and 1$M_{\oplus}$ can lose their captured hydrogen envelopes
during the saturation phase of their host stars
if the dust depletion factor $f$ is not much lower
then 0.01.

Fig. 5 shows the evolution of the masses of accumulated hydrogen for
heating efficiencies $\eta$ of 15\% and 40 \%, dust depletion
factors $f$ of 0.001, 0.01, 0.1, as well as relative accretion
rates $\dot{M}_{\rm acc}/M_{\rm pl}$ of $10^{-5}$ yr$^{-1}$ and $10^{-6}$ yr$^{-1}$.
The units of atmospheric mass are given in Earth ocean equivalent
amounts of hydrogen EO$_{\rm H}$ during the saturated XUV phase of a Sun-like G star for
protoplanets with 0.5$M_{\rm \oplus}$, 1$M_{\rm \oplus}$ and
2$M_{\rm \oplus}$. Our results show that large martian size
planetary embryos with masses $\leq$0.1$M_{\rm \oplus}$ lose their
captured hydrogen envelopes very fast. This is also in agreement
with the results of Erkaev et al. (2013b) who found that
nebula-captured hydrogen of a proto-Mars at 1.5 AU will be lost
within 0.1--7 Myr. Because our $0.1M_{\rm \oplus}$ protoplanet is
located at 1 AU where the XUV flux is much higher compared to the
martian orbit, the captured hydrogen envelopes are lost faster.
One can also see that a protoplanet with a mass of
$\sim$0.5$M_{\rm \oplus}$ will lose its captured hydrogen envelope
if the relative accretion rate $\dot{M}_{\rm acc}/M_{\rm pl}$ is
$\geq 10^{-6}$ yr$^{-1}$. However, for a lower $\dot{M}_{\rm acc}/M_{\rm pl}$ that
corresponds to luminosities of $\sim 3.7\times
10^{23}$--$3.7\times 10^{24}$ erg s$^{-1}$, a 0.5$M_{\rm \oplus}$
protolanet orbiting a Sun-like star at 1 AU may have a problem to
get rid of its captured hydrogen envelope during its life time.
The same can be said about protoplanets with masses of
$\sim$1$M_{\rm \oplus}$.

One can also see from Fig. 5 that planets with $\sim$1$M_{\rm \oplus}$
may have a problem to lose their captured hydrogen envelopes
during the XUV saturation phase of their host star if the dust
depletion factor $f$ in the nebula is $<$ 0.01. However, as it was
shown in Erkaev et al. (2013a), an Earth-like planet with a
hydrogen dominated upper atmosphere may lose between $\sim$4.5
EO$_{\rm H}$ ($\eta$=15\%) and $\sim$11 EO$_{\rm H}$ during its
whole life time after the XUV saturation phase. From this findings
one can conclude that protoplanets, that captured not much more
than $\sim$10 EO$_{\rm H}$ can lose their nebula based hydrogen
envelopes during their life times.

However, for the Earth there is
no evidence that it was surrounded by a dense hydrogen envelope
since the past 3.5 Gyr. Therefore, one can assume that the nebula
in the early Solar System had a dust depletion factor $f$ that was
not much lower than $\sim 0.01$ and that the relative accretion
rates $\dot{M}_{\rm acc}/M_{\rm pl}$ were most likely in the range
between $10^{-6}$--10$^{-5}$ yr$^{-1}$. Lower values for $f$ and
$\dot{M}_{\rm acc}/M_{\rm pl}$ would yield hydrogen envelopes that
could not be lost during Earth's history. Alternatively, early
Earth may not have been fully grown to its present mass and size
before the nebula dissipated, so that Earth's protoplanetary mass
was probably $<$1$M_{\rm \oplus}$. In such a case our results
indicate that a captured hydrogen envelope could have been lost
easier or faster. However, even if the dust depletion factor plays
a less important role for such a scenario, our study indicates
that the relative accretion rate $\dot{M}_{\rm acc}/M_{\rm pl}$
should not have been much lower than $\sim 10^{-6}$ yr$^{-1}$.

Table 5 summarizes the integrated losses $L_{\rm \Delta t}$ during
90 Myr in \% of atmospheric mass for all studied nebula scenarios, by using the
results given in Tables 1 to 3. Because it is beyond the scope
of this study to run full hydrodynamic simulations for all
$f_{\rm env}$ and related exact $z_0$, we assume that
$z_0=0.15R_{\rm \oplus}$ if $f_{\rm env}\leq 0.001$ for protoplanets
with core masses between 0.1$M_{\rm \oplus}$--2$M_{\rm \oplus}$,
$1R_{\rm \oplus}$ and $1.2R_{\rm \oplus}$ for 3$M_{\rm \oplus}$
and 5$M_{\rm \oplus}$ core masses, respectively. For more massive
hydrogen envelopes we apply a $z_0$ that is closest to the corresponding
$f_{\rm env}$ and its related hydrogen escape rates given in
Table 4.

One can see that core masses that
are $\leq$1$M_{\rm \oplus}$ can lose their captured hydrogen
envelopes for nebula dust depletion factors $f\sim$0.001--0.1 and
relative accretion rates $\dot{M}_{\rm acc}/M_{\rm pl}\sim
10^{-5}$ yr$^{-1}$. For relative accretion rates $\dot{M}_{\rm acc}/M_{\rm
pl}\sim 10^{-6}$ yr$^{-1}$ planets with $\sim$1$M_{\rm \oplus}$ will lose
the hydrogen envelopes only for dust depletion factors
$f\geq$0.01.

Another important result or our study is that more
massive `super-Earths' would capture too much nebula gas that can
not be lost during their whole life time. One can see from Fig. 5
and also from the results presented in Table 5, that for a lower mass
`super-Earths' with $\sim2M_{\rm \oplus}$ only certain
nebula properties, namely a dust
depletion factor $f\sim$0.1 and relative accretion rates
$\dot{M}_{\rm acc}/M_{\rm pl}$ in the order of $\sim 10^{-5}$ yr$^{-1}$ show
a visible effect in the loss during the XUV saturation phase.
Under such nebula and accretion properties the planet would not
capture too much hydrogen from the nebula. According to the
results of Erkaev et al. (2013a), who studied the loss of hydrogen
from a massive `super-Earth' with 10$M_{\rm \oplus}$ and 2$R_{\rm
\oplus}$ after the XUV saturation phase, depending on the assumed
heating efficiency $\eta$, it was found that such planets may lose
from $\sim$1.5 EO$_{\rm H}$ ($\eta$=15\%) up to $\sim$6.7 EO$_{\rm
H}$ ($\eta$=40\%) during 4.5 Gyr in an orbit around a Sun-like
star at 1 AU. From this study one can assume that a $\sim2M_{\rm
\oplus}$ `super-Earth' that originated in the above discussed
nebula conditions, may lose its hydrogen envelope during several
Gyr when the XUV flux decreased from the saturation level to that
of today's solar value. But for all the other nebula dissipation
scenarios `super-Earths' will capture too much hydrogen, which can
not be lost during their life time.

`Super-Earths' with masses that are $\geq$2$M_{\rm \oplus}$
(see our test planets with 3$M_{\rm \oplus}$ and 5$M_{\rm \oplus}$)
will not get rid of their captured hydrogen envelopes if they grow to
these masses within the nebula life time.
Thus, such a planet resembles more a mini-Neptune than
a terrestrial planet. These results are in agreement with the frequent recent discoveries
of low density `super-Earths', such as several planets in the
Kepler-11 system, Kepler-36c, Kepler-68b, Kepler-87c
and GJ 1214b (e.g., Lissauer et al. 2011; Lammer et al. 2013; Lammer 2013;
Ofir et al. 2013).

\section{IMPLICATIONS FOR HABITABILITY }
\label{hab} From the results of our study we find that the nebula
properties, protoplanetary growth-time, planetary mass, size and
the host stars radiation environment set the initial conditions
for planets that can evolve as the Earth-like class I habitats.
According to Lammer et al. (2009b), class I habitats resemble
Earth-like planets, which have a landmass, a liquid water ocean
and are geophysically active (e.g. plate tectonics, etc.) during
several Gyr. As can be seen from Table 5, as well as from Fig 5,
the Solar System planets, such as Venus and Earth, may have had
the right size and mass, so that they could lose their nebula
captured hydrogen envelopes perhaps during the first 100 Myr after
their origin, and most certainly during the first 500 Myr.

It should also be noted that a hydrogen envelope of a moderate
mass around early Earth may have acted during the first hundred
Myrs as a shield against atmosphere erosion by the solar wind
plasma (Kislyakova et al. 2013) and could have thus protected
heavier species, such as N$_2$ molecules, against rapid
atmospheric loss (Lichtenegger et al. 2010; Lammer et al., 2013b; Lammer 2013).
This idea was also suggested recently by Wordsworth \& Pierrehumbert
(2013) in a study related to a possible H$_2$-N$_2$
collision induced absorption greenhouse warming in Earth's early
atmosphere with low CO$_2$ contents so that the planet could sustain surface liquid water
throughout the early Sun's lower luminosity period, the so-called faint
young Sun paradox.

The before mentioned studies are quite relevant, because our results indicate also that there should be Earth-size and mass
planets discovered in the future that have originated in
planetary nebulae conditions where they may not have lost their
captured, nebula-based protoatmospheres. This tells
us that one should be careful in speculating about the evolution
scenarios of habitable planets, such as in the case of the
recently discovered $1.6R_{\oplus}$ and $1.4R_{\oplus}$
Kepler-62e and Kepler-62f `super-Earths' (Borucki et al. 2013),
especially if their masses and densities are unknown.

From our results shown in Table 5, it is not comprehensible why these authors
concluded that Kepler-62e and Kepler-62f should have lost their primordial
or outgassed hydrogen envelopes despite the lack of an accurately measured mass
for both `super-Earths'. If one assumes that both `super-Earths' with their
measured radii of $\sim$1.5$R_{\rm \oplus}$ have an Earth-like rocky
composition, one obtain masses of $\sim$3--4$M_{\rm \oplus}$ (e.g., Sotin et al. 007).
As one can see from Table 5, depending
on the nebula properties and accretion rates, `super-Earths' inside this mass
range can capture hydrogen envelopes containing between hundreds and
several thousands EO$_{\rm H}$. Although, Kepler-62 is a K-type star which
could have remained longer in the XUV-saturation than $\sim 100$ Myr, from
our model results it is obvious that, contrary to the assumptions by Borucki et al. (2013), the XUV-powered hydrodynamic escape rates are most likely not
sufficient for `super-Earths' within this mass range to get rid of
their hydrogen envelopes if they orbit inside the HZ of their host star.

In addition to the thermal escape, Kislyakova et al. (2013)
studied the stellar wind induced erosion of hydrogen coronae of
Earth-like and massive `super-Earth'-like planets in the habitable
zone and found that the loss rates of picked up H$^+$ ions within a wide
range of stellar wind plasma parameters are several times less
efficient compared to the thermal escape rates. This trend
continoued by the study of the stellar wind plasma interaction and
related H$^+$ ion pick up loss from five `super-Earths´ of the Kepler-11
system (Kislyakova et al. 2014). Thus, stellar wind
induced ion pick up does not enhance the total loss rate from
hydrogen dominated protoatmospheres and may not, therefore, modify
the losses given in our Tables 4 and 5 significantly.

\section{Conclusion}
\label{con} We modelled the capture of nebula-based hydrogen
envelopes and their escape from rocky protoplanetary cores of
`sub-' to `super-Earths' with masses between 0.1--5$M_{\rm
\oplus}$ in the G-star habitable zone at 1 AU. We found that
protoplanets with core masses that are $\leq$1M$_{\rm \oplus}$ can
lose their captured hydrogen envelopes during the active XUV
saturation phase of their host stars, while rocky cores within
the so-called `super-Earth' domain most likely can not get rid of
their nebula captured hydrogen envelopes during their whole lifetime.
Our results are in agreement with the suggestion that Solar System terrestrial
planets, such as Mercury, Venus, Earth and Mars, lost their
nebula-based protoatmospheres during the XUV activity saturation
phase of the young Sun.

Therefore, we suggest that
`rocky' habitable terrestrial planets, which can lose their
nebula-captured hydrogen envelopes and can keep their outgassed or
impact delivered secondary atmospheres in habitable zones of
G-type stars, have most likely core masses with 1$\pm$0.5$M_{\rm
\oplus}$ and corresponding radii between 0.8--1.15$R_{\rm \oplus}$.
Depending on nebula conditions, the formation
scenarios, and the nebula life time, there may be some planets
with masses that are larger than 1.5$M_{\rm \oplus}$ and lost
their protoatmospheres, but these objects may represent a minority
compared to planets in the Earth-mass domain.

We also conclude that several recently
discovered low density `super-Earths' with known radius and mass
even at closer orbital distances could not get rid of their
hydrogen envelopes. Furthermore, our results indicate that one
should expect many `super-Earths' to be discovered in the near
future inside habitable zones with hydrogen dominated atmospheres.
\begin{landscape}
\begin{table}
\renewcommand{\baselinestretch}{1}
\caption{Captured hydrogen envelopes in units of EO$_{\rm H}$ and
integrated losses $L_{\rm \Delta t}$ during 90 Myr in \%, during the G-star XUV
saturation phase with a 100 times higher XUV flux compared to that
of the present solar value for protoplanets
with core masses of 0.1$M_{\rm \oplus}$, 0.5$M_{\rm \oplus}$,
1$M_{\rm \oplus}$, 2$M_{\rm \oplus}$, 3$M_{\rm \oplus}$, and
5$M_{\rm \oplus}$. All losses are calculated for two heating
efficiencies $\eta$ of 15\% and 40\% and the three nebula
conditions described in Tables 1 to 3 for dust depletion factors
$f\sim$0.001--0.1.}
\begin{center}
\begin{tabular}{ccccccccccc}
     &                          &$f=0.001$,$\eta$=15-40\%   &   &  &    $f=0.01$,$\eta$=15-40\%         &     &  & $f=0.1$,$\eta$=15-40\%       &  \\\hline

$M_{\rm core}$ [$M_{\rm \oplus}$]&  $\frac{\dot{M}_{\rm acc}}{M_{\rm pl}}$ [yr$^{-1}$] & $M_{\rm atm}$ [EO$_{\rm H}$] & $z_0$ [$R_{\rm \oplus}$] & $L_{\rm \Delta t}$ [\%] &  $M_{\rm atm}$ [EO$_{\rm H}$] &$z_0$ [$R_{\rm \oplus}$] &  $L_{\rm \Delta t}$ [\%] &  $M_{\rm atm}$ [EO$_{\rm H}$] & $z_0$ [$R_{\rm \oplus}$] & $L_{\rm \Delta t}$ [\%]           \\\hline\hline
0.1  &              $10^{-5}$  &  0.014  & 0.15&   100         & 0.0022& 0.15 &100           & 0.00017& 0.15 & 100\\
0.5  &              $10^{-5}$  &  1.429  & 0.15& 100         & 0.227 & 0.15   & 100           & 0.029  & 0.15 &100\\
1    &              $10^{-5}$  &  9.608  & 0.15&8--31.2$\,^a$ & 1.682 & 0.15  & 100        & 0.316  & 0.15 & 100\\
2    &              $10^{-5}$  &  66.01  & 1   & 2.57--4.5& 13.84 & 0.15      & 12.3--21.7& 4.747  & 0.15 & 35.8--63.2$\,^a$\\
3    &              $10^{-5}$  &  211.5  & 1   &2.65--5.15& 52.86 & 1   & 10.6--20.6& 23.37  & 1 & 23.9--46.6\\
5    &              $10^{-5}$  &  956.2  & 1.5 &0.65--1.3& 299   & 1.2 & 2--4.1& 169.6  & 1.2 & 3.65--7.3\\
\hline
0.1  &              $10^{-6}$  &  0.047  & 0.15& 100         & 0.014 & 0.15 &  100          & 0.0023& 0.15 &100\\
0.5  &              $10^{-6}$  &  6.275  & 0.15& 62.15--100$\,^a$ & 1.776 & 0.15 &  100          & 0.0246& 0.15 & 100\\
1    &              $10^{-6}$  &  44.6  & 0.15 & 3.87--6.72  & 12.8  & 0.15 & 13.5--23.4& 2.313 & 0.15 & 74.8--100$\,^a$\\
2    &              $10^{-6}$  &  319  & 1     &  0.53--0.94& 96.14 & 0.15 & 1.77--3.12& 24.95 & 0.15 & 6.8--12\\
3    &              $10^{-6}$  &  1030 & 1     &0.54--1   & 326.2 & 1 & 1.7--3.34& 107.8 & 1  & 5.2--10.1\\
5    &              $10^{-6}$  &  4653 & 1.5   & 0.13--0.35& 1600  & 1.5 & 0.38--0.78& 690.8 & 1.2 & 0.9--1.8\\
\hline
0.1  &              $10^{-7}$  &  0.150  & 0.15 & 100              & 0.052 & 0.15 & 100        & 0.014& 0.15 & 100 \\
0.5  &              $10^{-7}$  &  28.2   & 0.15 & 13.8--33   & 8.673 & 0.15 & 45--100$\,^a$    & 1.971& 0.15 &100 \\
1    &              $10^{-7}$  &  210.0  & 1    & 0.82--1.42 & 65.75 & 0.15 & 2.63--4.56       & 16.8 & 0.15 & 10.3--55.35\\
2    &              $10^{-7}$  &  1565   & 1    & 0.1--0.2   & 500.3 & 1 & 0.34--0.6& 153.2           & 0.15 &1.2--1.9\\
3    &              $10^{-7}$  &  5270   & 1.35 & 0.1--0.3   & 1680  & 1.35 & 0.47--0.92&584.1        & 1.35 &0.95--1.86\\
5    &              $10^{-7}$  &  28620  & 4    & 0.02--0.25 & 8117  & 4 &  0.09--0.2& 3250           & 1.5 & 0.24--0.49\\
\hline
0.1  &              $10^{-8}$  &  0.392  & 0.15 & 100       & 0.183  & 0.15 &  100       & 0.054 & 0.15 & 100 \\
0.5  &              $10^{-8}$  &  116.7  & 0.5  & 9.4--20.4 & 41.95  & 0.15 &  26.2--56.7& 10.75 & 0.15 & 100\\
1    &              $10^{-8}$  &  1002   & 1    &    1.73--4& 332.6  & 1    & 5.2--12.1  & 93.2  & 0.15 & 1.85--3.2\\
2    &              $10^{-8}$  &  9934   & 4    & 1.93--4   & 2670   & 1    & 0.23--0.43 & 826.1 & 1    & 0.72--1.39\\
3    &              $10^{-8}$  &  56640  & 4    & 0.19--0.46& 9797   & 4    & 1.1--2.65  & 3066  & 1.35 & 0.26--0.5\\
5    &              $10^{-8}$  &  105816 & 4    & 0.03--0.07& 103725 & 4    & 0.03--0.072& 17810 & 4    & 0.23--0.42\\
\hline
\end{tabular}
\end{center}
$^a)$ Protoplanets that may lose their remaining hydrogen remnants
during the rest of their lifetime after the XUV activity saturation phase ends
and the solar/stellar XUV flux begins to decrease (see Sect. 2) to $\leq$
the present time solar value at 1 AU (Erkaev et al. 2013a).
\normalsize
\end{table}
\end{landscape}
\section*{ACKNOWLEDGMENTS}
The authors acknowledge the support by the FWF NFN project
S11601-N16 `Pathways to Habitability: From Disks to Active Stars,
Planets and Life', and the related FWF NFN subprojects, S 116
02-N16 `Hydrodynamics in Young Star-Disk Systems', S116 604-N16
'Radiation \& Wind Evolution from T Tauri Phase to ZAMS and
Beyond', and S116607-N16 `Particle/Radiative Interactions with
Upper Atmospheres of Planetary Bodies Under Extreme Stellar
Conditions'. K. G. Kislyakova, Yu. N. Kulikov, H. Lammer, \& P.
Odert thank also the Helmholtz Alliance project `Planetary
Evolution and Life'. M. Leitzinger and P. Odert acknowledges
support from the FWF project P22950-N16. N. V. Erkaev acknowledges
support by the RFBR grant No 12-05-00152-a. Finally, the authors
thank the International Space Science Institute (ISSI) in Bern,
and the ISSI team `Characterizing stellar- and exoplanetary
environments'.

\end{document}